\documentclass[12pt,preprint]{aastex}
\def\lax {\ifmmode{_<\atop^{\sim}}\else{${_<\atop^{\sim}}$}\fi}  
\def\gax {\ifmmode{_>\atop^{\sim}}\else{${_>\atop^{\sim}}$}\fi}  
\def\gtorder{\mathrel{\raise.3ex\hbox{$>$}\mkern-14mu
             \lower0.6ex\hbox{$\sim$}}}

\begin{document}

\title{GRO J1655-40: Early Stages of the 2005 Outburst}

\author{ N. Shaposhnikov\altaffilmark{1,2}, J. 
Swank \altaffilmark{1}, C.R. Shrader \altaffilmark{1,2},  
M. Rupen\altaffilmark{3}, V. Beckmann \altaffilmark{1,4},
 C.B.~Markwardt\altaffilmark{1,5} and D.A. Smith\altaffilmark{6}}

\altaffiltext{1}{ NASA Goddard Space Flight Center, Exploration of the 
Universe Division Greenbelt MD 20771; nikolai@milkyway.gsfc.nasa.gov}

\altaffiltext{2}{Universities Space Research Association, 10211 Wincopin
Circle, Columbia, MD 21044, USA }

\altaffiltext{3}{National Optical Astronomy Observatory,950 North Cherry 
Avenue, Tucson, AZ 85719, USA }

\altaffiltext{4}{Joint Center for Astrophysics, Department of Physics,
University of Maryland, Baltimore County, MD 21250, USA}

\altaffiltext{5}{Department of Astronomy, University of Maryland, College
Park, MD 20742, USA}

\altaffiltext{6}{Department of Physics, Guilford College, Greensboro, NC 27410, USA}

\begin{abstract}

The black-hole X-ray binary transient GRO J1655-40 underwent an outburst 
beginning in early 2005. We present the results of our multi-wavelength 
observational campaign to study the early outburst spectral and temporal 
evolution, which combines data from X-ray ({\it RXTE}, {\it INTEGRAL}),
radio ({\it VLA}) and optical ({\it ROTSE}, {\it SMARTS}) instruments.
During the reported period the source left quiescence and went through  four 
major accreting black hole states: low-hard, hard intermediate, 
soft intermediate and high-soft.
We investigated 
dipping behavior in the {\it RXTE} band and compare our results to the 1996--1997 
case, when the source was predominantly in the high-soft state, finding
significant differences. We 
consider the evolution of the low frequency quasi-periodic oscillations 
and find that the 
frequency strongly correlates with the spectral characteristics, 
before shutting off prior to the 
transition to the high-soft state. We model the broad-band high-energy 
spectrum in the context of empirical models, as well as more 
physically motivated thermal and bulk-motion Comptonization and Compton
reflection models. 
{\it RXTE} and {\it INTEGRAL} data together support a statistically significant 
high energy cut-off in the energy 
spectrum at $\approx 100-200$ keV during the low-hard state.
The {\it RXTE} data alone also show it very significantly during the transition, 
but cannot see one in the high-soft state spectra. 
We consider radio, optical and X-ray
connections in the context of possible synchrotron and synchrotron self-Compton
origins of X-ray emission in low-hard and intermediate states. In this outburst
of GRO J1655--40, the radio flux does not rise strongly with the X-ray flux. 

\end{abstract}


\section{Introduction}

X-ray Novae, also known as soft X-ray transients, are a subclass of the low-mass 
X-ray  binaries (LMXBs) for which prolonged periods of quiescence 
are occasionally interrupted by dramatic, accretion-powered  optical, UV, and 
X-ray outbursts \citep{csl97,mr04}. A majority of these binaries have been 
determined to contain compact primaries with mass functions exceeding the 
nominal 3-solar-mass limit for neutron stars; indeed, most of the known 
stellar mass black holes  are associated with this class of objects. These 
outbursts are frequently accompanied by radio emission which is generally 
associated with collimated outflows. The radio emission is transient in 
nature, and a wide range of behavior has been documented.  

Conventional classification of black hole (BH) spectral states includes  
five states, namely: {\it quiescent state}, {\it low-hard state}
 (LHS), {\it intermediate state} (IS), {\it high-soft} (or {\it thermally dominant}) 
{\it state} (HSS) and {\it very high} (or {\it steep power law}) {\it state}. \citet{mr04} 
recently reviewed these states. They advocate the names in parentheses, as being more accurately 
descriptive. However, we follow tradition in the use of the abbreviation HSS.  
The LHS energy spectrum is dominated by a hard (spectral index 1.4--1.6)
 power law. A thermal component is either very small or not seen at all.
Strong variability (up to 40 \%) in the form of band-limited noise 
is observed in this state, often accompanied by 
quasi-periodic oscillations (QPO) seen in the Fourier Power Density Spectrum (PDS)
 as narrow peaks. Stable radio emission with a flat spectrum  has been 
observed during the LHSs of a set of BH X-ray sources. 
It has been associated with steady relativistic jets and outflows
\citep{gallo}.
BH transients are usually observed in the LHS when their luminosity is
less than 5 \% of the Eddington luminosity. The HSS, in contrast to the LHS, 
is dominated by a thermal
component with a characteristic temperature of $\approx$ 0.5--1.0 keV, which is 
attributed to bright emission from an optically thick accretion disk. In the 
HSS the power law
index is steeper ($>$2.0) and variability is suppressed to not more than several
percent (which is dominated by red noise, i.e. $P(f)\sim 1/f$). The LHS and the HSS are the 
most well observed and documented BH states.  During transitions 
between them, a BH source enters an IS, where the source may exhibit diverse behavior
with mixed LHS and HSS properties. 
\citet{hb05} considered the phenomenology of the IS in detail and introduced two substates,
which they call {\it hard intermediate state} (HIMS) and  
{\it soft intermediate state} (SIMS) \citep[See also][]{bel05,bel06}.
For the HIMS the energy spectrum is softer than in
the LHS and the thermal component is more pronounced. The PDS is still dominated by
band-limited noise with characteristic frequencies higher than those observed for the LHS.
Radio emission is also observed in the HIMS, but with a slightly steeper spectrum.
In the SIMS the disk component starts to dominate the energy spectrum and 
further softening is observed. The total rms variability drops sharply.
The PDS is a sum of band-limited and power law components, or is sometimes even more
complicated in shape. This subdivision of IS into HIMS and SIMS 
is consistent with the behavior of GRO J1655-40 during the hard-to-soft transition that we observed,
so we adopt this terminology when we want to refer to a particular intermediate substate.
In our data we identify LHS, HIMS, SIMS, and HSS periods.
 
The X-ray binary 
GRO~J1655--40 is a well known example of BH X-ray transient, having 
undergone several major outbursts within the the last 12 years.
It was discovered by the BATSE instrument onboard the 
Compton Gamma-Ray Observatory in mid 1994 \citep{zh94,har95}.
The secondary star being relatively bright, the
binary parameters are exceptionally well 
determined amongst LMXBs. 
%\citet{OB97}, using  
The most recent optical  photometry of \citet{greene01} led to 
 a BH mass estimate of $6.3\pm0.5 M_\odot$. \citet{hr95}
inferred a distance to GRO~J1655--40 of $3.2\pm0.2$ kpc from radio jet analysis.
 \citet{mir02} discussed in detail different methods of distance estimates
and concluded that for GRO~J1655--40 the upper limit is 3.5 kpc and the lower
limit is 0.9 kpc. 
%GRO~J1655--40 was discovered by the BATSE instrument onboard the 
%Compton Gamma-Ray Observatory in mid 1994 \citep{zh94,har95}.
The light curve of the discovery outburst was irregular, 
in the sense that it deviated 
substantially from the often seen "fast-rise exponential decay" form 
\citep{csl97}.  It was instead characterized by several distinct peaks, with 
subsequent sporadic, lower amplitude outburst activity continuing into 1995. 
These initial events were correlated with the contemporaneous 
radio observations. Most notable was the discovery of 
radio jets which exhibited apparent superluminal motion \citep{har95,hr95,tin95}. 
The later, low-amplitude events however, 
were apparently not associated with additional plasma ejections leading to 
the type of radio emission initially seen \citep[e.g.][]{tav96}.   In April 
1996, after a short period of quiescence, a new outburst appeared in the data 
of the {\it Rossi X-ray Timing Explorer} ({\it RXTE})
All-Sky Monitor \citep{rem96}. This event also consisted of 
two giant flares which lasted 100 and 222  days. Although the radio behavior was far less 
dramatic, extensive 
{\it RXTE} coverage allowed this outburst to be studied in greater detail 
than the 1994--1995 events. However, pointed {\it RXTE} observations started
when the source was already in the HSS and the state transition was
not covered. The 2005 outburst was recognized at a very early stage and 
observers at different wavelengths were relatively prepared and successful at 
responding relatively quickly. 
In this paper we present the analysis of the first multi-frequency
observations of an initial  hard-to-soft state transition for a GRO~J1655--40 outburst. 
The data 
comprise one the most detailed multi-frequency observations of the initial 
hard-to-soft transition in a BH transient [see \citet{broc02} for a review 
of the observed state transitions in BH sources].

On February 17, 2005 (MJD 53419) {\it RXTE} Proportional Counter Array (PCA) bulge scan observations
detected emission consistent with coming from the position of 
GRO~J1655--40 and strongly suggested a new outburst \citep{mar05}. The 
flux in the 2-10 keV band was 4.1 +/- 0.6 mCrab. 
Near infra-red (NIR) J photometry on February 20.39 UT found the source 0.5 mag 
brighter than at its brightest in quiescence \citep{atel417} and the NIR activity was confirmed
on February 21.3-21.4 \citep{atel418}. 
NRAO {\it Very Large Array Radio Observatory} ({\it VLA})
observations on February 20.6 found that it was again a detectable
radio source \citep{atel419}. A campaign of daily pointed {\it RXTE} observations
started on February 21, 2005 (MJD 53423).  These observations and the PCA bulge scans 
triggered our  program of Target of Opportunity (ToO) observations with 
INTernational Gamma-Ray Astrophysical Laboratory ({\it INTEGRAL}), the goal 
of which was to characterize the early-rise phase of X-ray nova events. 
Subsequently the source brightened and triggered other programs for observing bright 
transients.  The observations 
we consider in this paper span a 25 day period starting on 21 February 2005. We have 
utilized the radio flux-history from an ongoing program of transient 
monitoring at the {\it VLA}. The outburst was also monitored by the {\it Swift}
observatory with a series of pointed observations \citep{broc06}, the
first of which fell on the initial LHS within the time frame of the data
 reported here. The source was observed extensively until the outburst faded, by  {\it RXTE}
\citep{ho05a,ho05b,ho05c} and other missions.

In this paper we report on the results of our 
multi-wavelength observational campaign, beginning with the outburst detection 
and ending on March 16, 2005 (MJD 53445) after the source completed 
the hard-to-soft state transition. The data are described in \S 2. 
Our analysis of the {\it INTEGRAL} and {\it RXTE} observations is presented in \S 3.
This includes light curve analysis as well as our spectral model fitting. We discuss 
dipping behavior in \S 4. Interpretation and conclusion are presented in \S 5. 

\section{Observations and Data Analysis}

Our data combine  programs of target of opportunity as well as publicly available  observations from 
{\it RXTE},
{\it INTEGRAL}, 
the {\it VLA}, 
the {\it Robotic Optical transient Search Experiment} ({\it ROTSE}) 
and the {\it Small and Moderate Aperture Research 
Telescope System}\footnote{http://www.astro.yale.edu/smarts/}({\it SMARTS}. 
Below we provide details for each data set.

\subsection{{\it RXTE}}

{\it RXTE}'s campaign of pointed observations 
of the new outburst from GRO~J1655--40 started on February 21, 2005 (MJD 53422)
and provided data on almost a  daily basis. Observations were taken under the 
following {\it RXTE} Proposals: 90058, 90428, 90404, 90704 and 91702. 
In addition to Standard PCA data modes, 
high time and energy resolution data were collected.  For low and moderate 
count rates an Event mode with $\sim125$ microsecond time 
resolution in 64 energy bins was collected. For high source fluxes the high 
resolution data were provided by two Single Bit Modes covering PCA channels
 0-13 and 14-35  respectively, along with an Event mode  for counts 
registered in channels above 36. 
 
Data from the deep monitoring proposal 91702 were made publicly
available and Homan et al. (2005) also posted energy and power
spectra, as well as light curves. These observations began on March 7, 2005 (MJD 53436) shortly
after the source started the transition from the LHS. We have
used data from some of these observations to put the LHS
and the IS in the context of the developing outburst. We
reduced and analyzed them ourselves together with the previous
data for a uniform treatment.
 
The data reduction and analysis was performed using FTOOLS 5.3 software. 
For spectral analysis we use the PCA Standard2 data mode and the Standard Archive 
HEXTE Mode.  Standard dead time correction was applied to all spectra. 
Spectra were modeled using the XSPEC11.0 astrophysical fitting package.
We used 3.0-30.0 keV and 18.0-300.0 keV energy intervals for
PCA and HEXTE data correspondingly. We added  1\% systematic error to the data during our spectral 
analysis. The uncertainties on spectral model
parameters were calculated using 1$\sigma$ confidence intervals.
For Fourier Analysis we use high time resolution PCA data modes, 
combining counts from different modes to get a signal from the entire PCA 
energy range.  After rebinning the data to obtain the Nyquist frequency value of 
1024 Hz we calculate individual power density spectra (PDS) for consecutive 
128 second intervals and averaged  them for each {\it RXTE} orbit. 
For timing analysis we utilize our own IDL Library routines. 

\subsection{{\it INTEGRAL}}

An {\it INTEGRAL} Target of Opportunity (ToO) observation of GRO~J1655--40,
previously approved as part of the General Program \citep[see][]{wink03}, was
performed between MJD 53425.21 and MJD 53427.60. 
%The {\it INTEGRAL}
%data are shown on the bottom panel  of Figure \ref{lc}, where we plot {\it INTEGRAL}/ISGRI
% and HEXTE/Cluster A light-curves. 
The observations consist of a series of dither pointings lasting between 58
and 146 minutes each. In our case a total of 49 pointings were included in
the analysis. Data from the {\it INTEGRAL} hard X-ray instruments
IBIS/ISGRI and SPI, as well as from the $3 - 35 \rm \, keV$ X-ray monitor
JEM-X were used in our analysis. All three instruments use the coded mask
imaging technique.
Data reduction was performed using the standard OSA 5.0 analysis software
package available from the {\it INTEGRAL} Science Data Centre \citep{cou03}.
Note that because of the nature of coded mask imaging, the whole
sky image taken by the instrument has to be taken into account in the
analysis, as all sources in the field of view contribute to the background
\citep{car87}. The total exposure time of 178 ks is the ISGRI
effective on-source time. This value is approximately the same for the
spectrograph SPI (186 ks), but the JEM-X monitor (17.5 ks)  covers a much
smaller sky area. Thus in the case of dithering observations, the source is
not always in the field of view of the monitor.
 
In order to achieve more complete time coverage, we added IBIS/ISGRI data of
GRO J1655--40 provided by {\it INTEGRAL}'s Galactic Bulge monitoring
program (PI E. Kuulkers). Within this program GRO J1655--40 was observed
once every orbit (i.e. every three days) for 12.6 ks \citep{erik06}.
                                                                                
\begin{figure}[ptbptbptb]
\includegraphics[scale=0.7,angle=0]{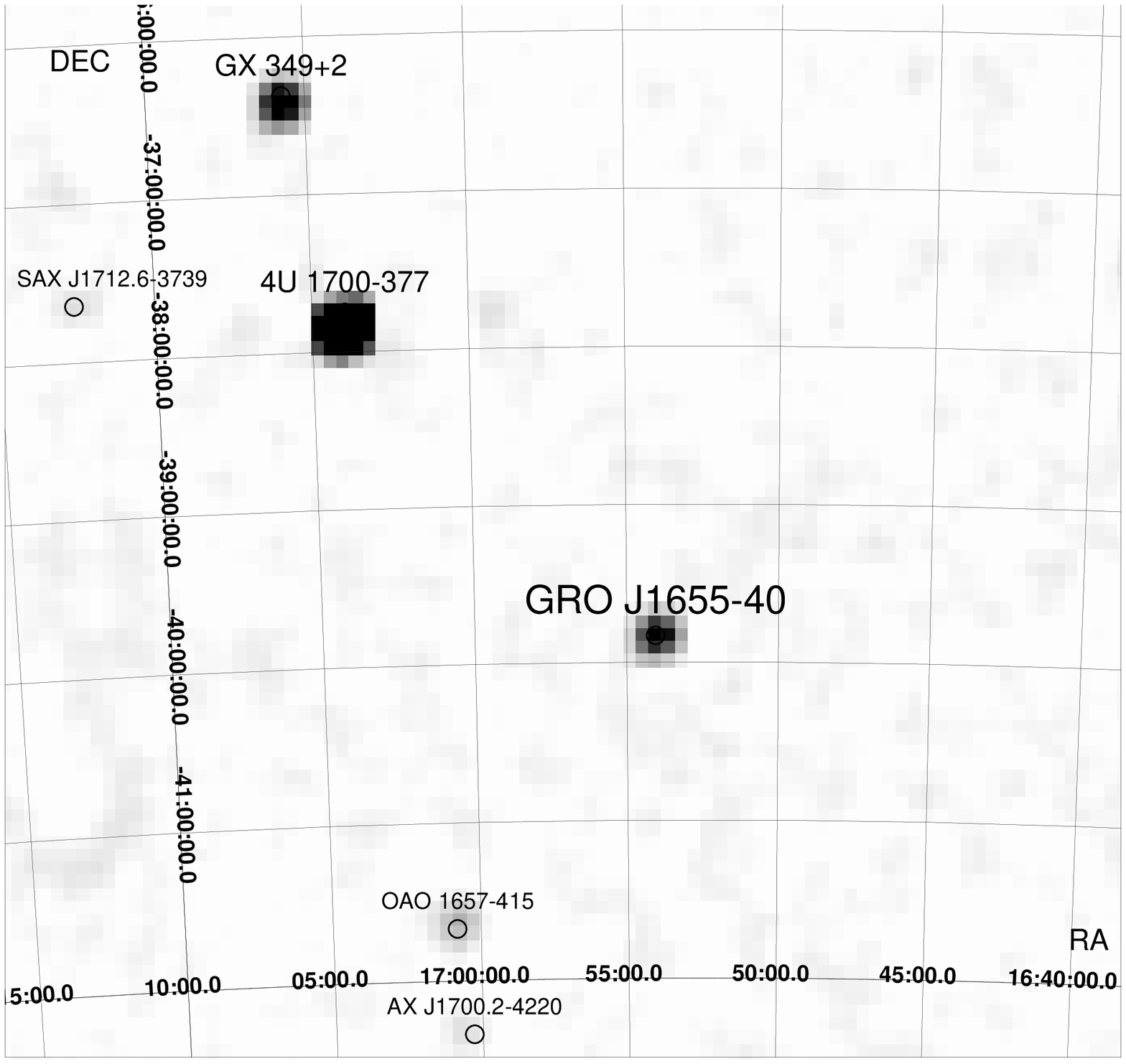}
\caption{{\it INTEGRAL} IBIS/ISGRI 20 -- 40 keV significance map, based on 178 ks
exposure time. No nearby sources are detected and no source confusion
affects the data of GRO J1655--40. In this data set the average 20-60 keV flux of GRO J1655-40 is
$9.64\pm 0.05$ counts s$^{-1}$ or $(8.80\pm 0.05)\times 10^{-10}$ ergs cm$^{-2}$ s$^{-1}$. 
The $5\sigma$ detection limit 
during this observation is $2.3\times 10^{-11}$  ergs cm$^{-2}$ s$^{-1}$.}
\label{integral_map}
\end{figure}

The analysis of the ISGRI data is based on a cross-correlation procedure
between the recorded image on the detector plane and a decoding array
derived from the mask pattern \citep{gold03}. Imaging analysis led
to a detection significance of $93 \sigma$, $5 \sigma$, and $20 \sigma$,
for the ISGRI (20 -- 200 keV), SPI (20 -- 200 keV), and JEM-X (3 -- 35
keV) data, respectively.The {\it INTEGRAL} IBIS/ISGRI 20 -- 40 keV significance map is 
shown on Figure \ref{integral_map}. No nearby point sources are detected closer
than 1$^\circ$ ({\it RXTE} Collimator FWHM) and brighter than a mCrab above 20 keV.

\subsection{{\it VLA}}

The radio data are presented in Table \ref{vladata}.
The data were taken with the {\it VLA} in its `B'
configuration, using standard continuum modes to obtain two 50\,MHz IF pairs of
right- and left-circular polarization, for a total bandwidth of 200\,MHz.
Flux densities were derived from 3C\,286, which was also observed during
most runs.  Phases were calibrated using interleaved observations of strong
nearby point sources.  The data were reduced  using the 31DEC04 version
of the National Radio Astronomy Observatory's (NRAO's) AIPS package \citep[e.g.][]
{nrao-aips}.

Observations were made at 1425, 4860, 8460 and 22460 \,MHz.
No extension is seen in any of the images, for which the typical
resolution was $5.1\times 1.2\rm\,(4.86\,GHz/\nu)$\,arc-seconds, oriented
roughly north-south. Gaussian fits to each image showed no convincing elongations, with
typical upper limits of 0.8 to 1 arc-seconds at 8.46\,GHz.
The 1-sigma flux uncertainties are listed in column 6 of Table 1. From these values it is seen that absolute flux calibration 
is accurate to typically 10-15\%.  Upper limits in the figures are the sum of the nominal flux
density at the position of the source, plus three times the rms noise.

While the observed position of the source shifts significantly from
epoch to epoch, this is probably simply due to variable atmospheric effects
at the low elevations required at the {\it VLA} to observe such a southern source.
Indeed, shifts of similar magnitude are seen between the different observing
frequencies on a given day, and other sources in the field at 1.4 and
4.9\,GHz also move around from epoch to epoch.  Minimizing
the effects of outliers, and removing two observations which were clearly
unreliable, we obtain a mean (J2000) position in arcseconds of
\[\alpha = \phn16^h\ 54^m\ 00^s.139 \pm 0.008,\,
  \delta = -39^\circ\ 50^{'}\ 44^{''}.7  \pm 0.2\]
where the error bars were derived after adding 0.5\,arcsec in quadrature to 
the statistical errors found for each individual image, as required to
give an average L1 deviation of $1\sigma$ from this mean.
Within errors this position is consistent with the optical position that \citet{bailyn95}
found in 1994. The measurements are not accurate enough to detect the proper
motion of 5.2 mas yr$^{-1}$ derived by \citet{mir02}.

\subsection{Optical Data}

We use data from {\it ROTSE} \citep[{\it ROTSE}, ][]{rotse} 
 for the optical light-curve \citep{smith05}. {\it ROTSE} images are unfiltered, 
calibrated to the USNO A2.0 R band, and then
converted to flux units via numerical integration of the CCD response function.

From {\it SMARTS} 
we use the data obtained by \citet{atel418} in B, V, I, J and K bands 
with the ANDICAM instrument on the SMARTS 1.3m telescope 
at Cerro Tololo Inter-American Observatory on February 21 (MJD 53422.3-53422.4). 
Optical magnitudes 
at quiescence were taken from \citet{greene01}. 
The latter's uncertainty in period would imply the photometric 
phase was within about 0.1 of 0.97.

 \section{Results}
 
\subsection{Multi-frequency Evolution}
 
\begin{figure}[ptbptbptb]
\includegraphics[scale=0.7,angle=0]{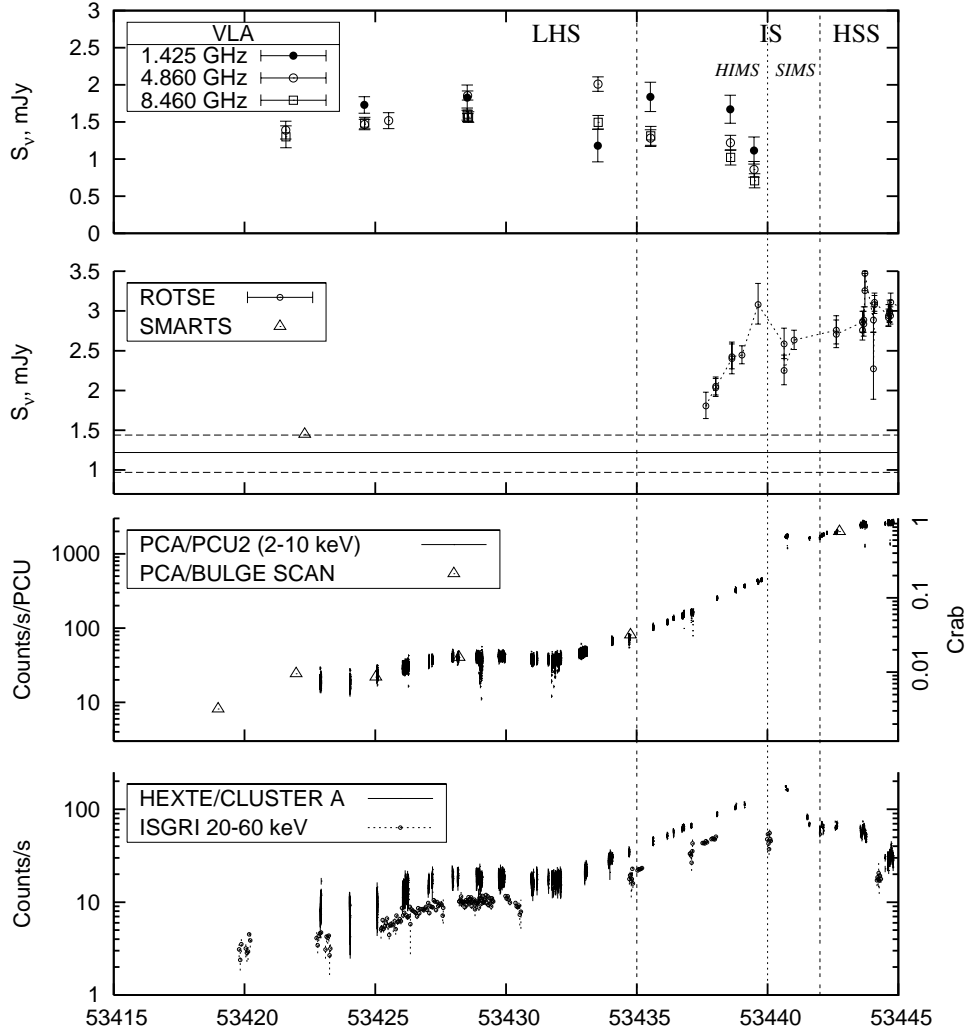}
\caption{
Radio ({\it VLA}), R-band optical ({\it SMARTS},{\it ROTSE}) and X-ray({\it RXTE},{\it INTEGRAL}) light-curves are presented on 
the first three panels (from the top downward). 
To obtain R band flux from {\it SMARTS} data we linearly interpolated the data from 
\citet{atel418}. Vertical lines indicate the approximate dates when the source changes
its spectral state. The solid horizontal line on the panel with the optical data corresponds to
the quiescent {\it ROTSE} flux. The dashed lines are the minimum and maximum of quiescent 
ellipsoidal variations of the companion.
The ephemeris during this time is uncertain by 0.27 d. The ISGRI light curve comprises
both the {\it INTEGRAL} bulge scan and our proposal data.
 }
\label{lc}
\end{figure}

\begin{figure}[ptbptbptb]
\includegraphics[scale=0.7,angle=0]{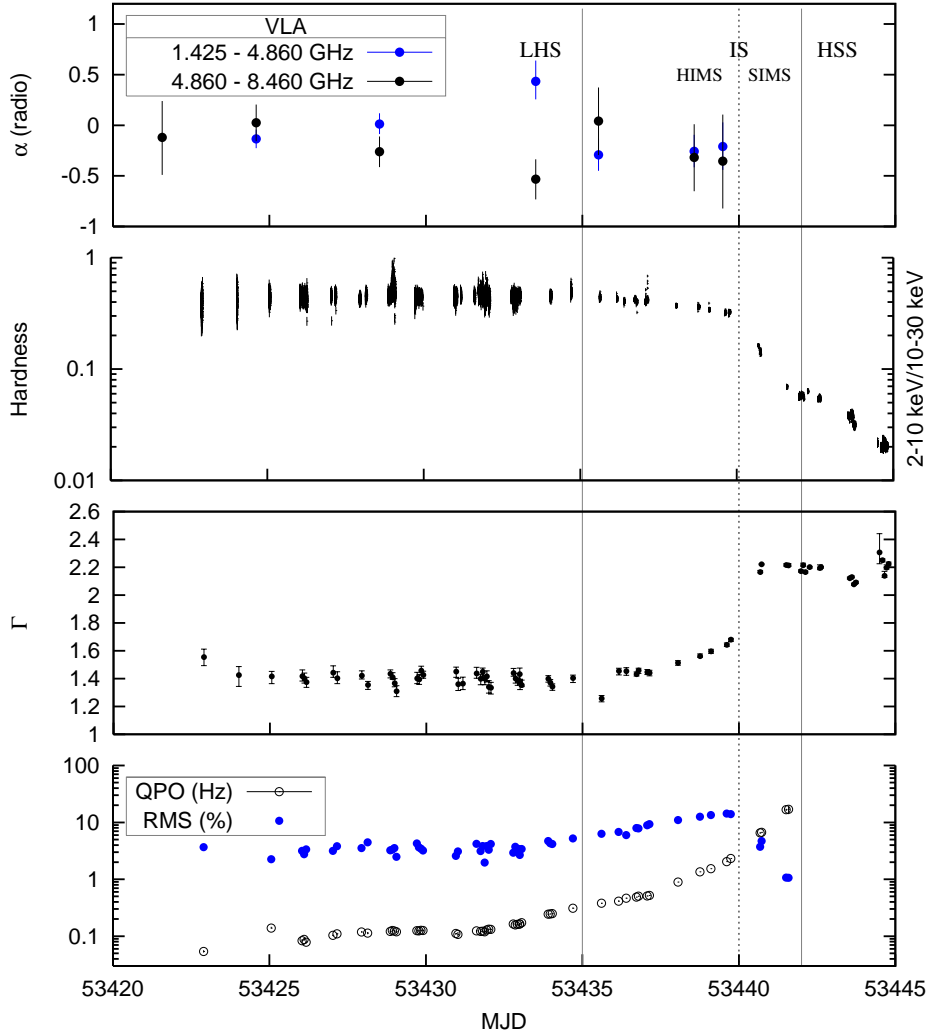}

\caption{Quantities, derived from the data (from the top to the bottom): spectral index,
calculated from radio data using pairs of observational points (1.425 and 4.860 GHz - red,4.860 and 8.460 GHz - black);
 PCA hardness ratio; X-ray spectral index obtained from the {\it power law + black body} 
model fit to the PCA data; QPO frequency and its rms variability.  
Important changes occur in the source behavior around MJD 53440,
when the radio flux drops effectively to zero (see Figure \ref{lc}), while the QPO amplitude 
drops down abruptly, to disappear two days later. Vertical lines mark the boundaries
between spectral states which are discussed in the text. }
\label{derived}
\end{figure}

The radio, optical, and X-ray light curves are shown on Figure 
\ref{lc}. In Figure \ref{derived} we plot several interesting properties
of the source that we derived from the data.
  The rising stage of the outburst consists of three 
distinct phases. Specifically, after an initial 5 day rise from $\sim8.5$ to 
$\sim17$ mCrab (2--10 keV), the source dwelt for $\sim4$ days in 
the LHS. On March 4 (MJD 53433) 
 the flux started rising again, accompanied by a gradually softening spectrum indicating that 
the source entered the hard-to-soft transition phase. We follow the spectrum 
evolution for a month beginning from the start of {\it RXTE} pointed observations, when 
the source was in low-hard state (LHS), to the point where the source 
completes the transition to the high-soft state (HSS). We divided the  
period of interest into the three intervals, according to the source state.
We also subdivide the IS into two substates according to the \citet{hb05} classification.
On Figure \ref{sp_pds} we show representative energy and power spectra for
each interval.

\subsubsection{Low-Hard State (MJD 53418 to 53435)} 

The LHS stage lasted from the detection on February 17 (MJD 53418) through
March 6 (MJD 53435). After an initial 10 day rise with no detected
spectral change, the source entered a rather stable period that lasted for 
approximately 5 days with roughly constant X-ray
flux and hardness. The X-ray spectrum is well-fit by a power law of index $\approx$1.5,
while less than 10 \% of the emission is in a thermal component.
The PDS is represented by a broken power law typical for the LHS (also known as band-limited noise).
The lower frequency part is usually flat, while after a brake the slope is 1--1.5.
The low frequency QPO shows above this continuum as a narrow peak (see Figure 
\ref{sp_pds}, panel A). The luminosity at this juncture was about 0.5\% Eddington 
(for the distance of 3.2 kpc and the mass of 6.3$M_\odot$). The flat spectrum radio source is 
roughly steady at $\sim1.5\rm\,mJy$. \citet{atel418} and \citet{atel417} concluded the optical 
magnitudes were within the amplitude range of quiescent
ellipsoidal variations, while the infrared (J and K) were linearly elevated
but again roughly steady during this period. A high energy cut-off is observed (but sometimes 
with only marginal significance) in the X-ray energy spectrum. On March 3 (MJD 53432) the source 
entered a phase of exponential rise. However, the HIMS, designated by a start of softening 
(see Figure \ref{derived}), only starts three days later. 

\begin{figure}[ptbptbptb]
\includegraphics[scale=0.6,angle=-90]{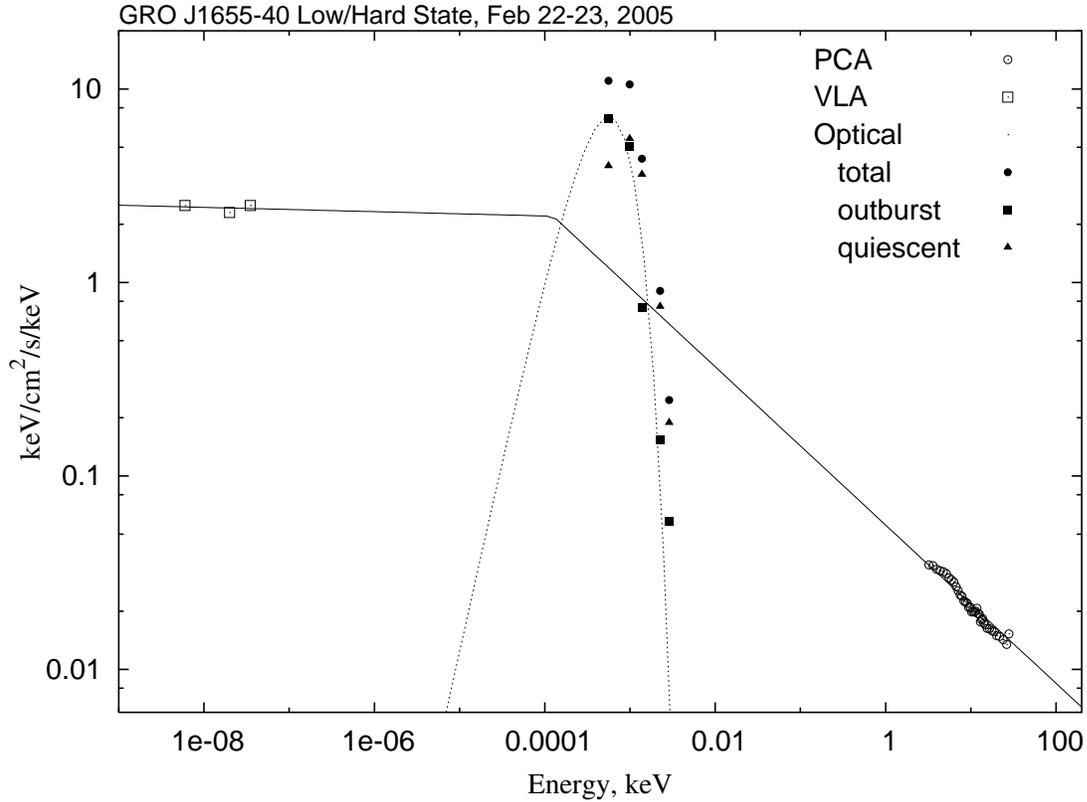}
\caption{
Multi-frequency energy spectrum of LHS (Feb 21, 2005) from
{\it RXTE}, {\it SMARTS} and {\it VLA}. 
Extrapolations of a  broken power law can join the radio and X-ray
data, with the break in the IR range (solid line), if absorption hides the UV. 
The outburst optical and infrared fluxes are calculated from the {\it SMARTS} results
\citep{atel418}. They suggest a black body shape with
$kT\sim 0.2$ eV, presumably coming from the outer disk (dashed line).}
\label{multi_spec}
\end{figure}

We combined radio, optical and X-ray data to construct 
the multi-frequency energy spectrum of the LHS shown on Figure \ref{multi_spec}.
The data are quasi-simultaneous in a sense that we combine radio 
and X-ray data closest in time to  the optical observations and the time offset between
observations in different wavelengths does not exceed 24 hours.
The overall spectrum is similar to the broadband LHS spectra for 
four other BH sources presented by \citet{gallo}. 
The radio shows a rather flat spectrum such that if it were extrapolated to higher frequencies
as a power law and the (unabsorbed) X-rays were extrapolated to lower frequencies as a power law, 
they would meet in the near infra-red range at $\sim0.1$ eV. 
This is similar to the IR break for GX 339-4
\citep{nowak05}. The optical data have an approximately 
black-body shape with $kT\sim 0.2$ eV.
On Figure \ref{multi_spec} we plot the total observed optical fluxes,
the average quiescent levels, and  outburst fluxes
calculated as the total minus the average quiescent flux.  
The low temperature of this component suggests that the outer accretion 
disk is its origin. Using the average 
quiescent fluxes should give an underestimate of the fluxes and 
an overestimate of kT, if the phase was indeed
near the shallower V band minimum.

\subsubsection{Intermediate State (MJD 53435 to 53442)}

\subsubsubsection{Hard Intermediate State}

The period from MJD 53435 through MJD 53440 is marked
by a roughly {\it exponential rise} in the X-ray 
flux, with the X-rays gradually softening.  The 2--10\,keV
flux rises by over an order of magnitude during this interval. 
Also a rise in the optical flux was observed with {\it ROTSE}.
Power law fits to the X-ray and optical data sets 
give characteristic rise times of 3--4 days, with the rise being slowest in
the optical and hard X-rays. The rise of the 2--10\,keV flux 
reflects the gradual softening of the source, seen also in
the power law index $\Gamma$ (Figure \ref{lc}) .
The radio data  are sparse, but they indicate a turn-over in the source flux and a 
steepening of the  spectrum. There is no sign of the
factor $\sim5$ increase like that of the simple $S_\nu\propto F_x^{0.7}$
radio-to-X-ray scaling observed for some black hole transients rising to the HSS 
transition \citet{gallo}. For this outburst of GRO~J1655--40, at least, the radio
emission is quenched relatively early in the evolution.
 
\begin{figure}[ptbptbptb]
\includegraphics[scale=1.0,angle=-90]{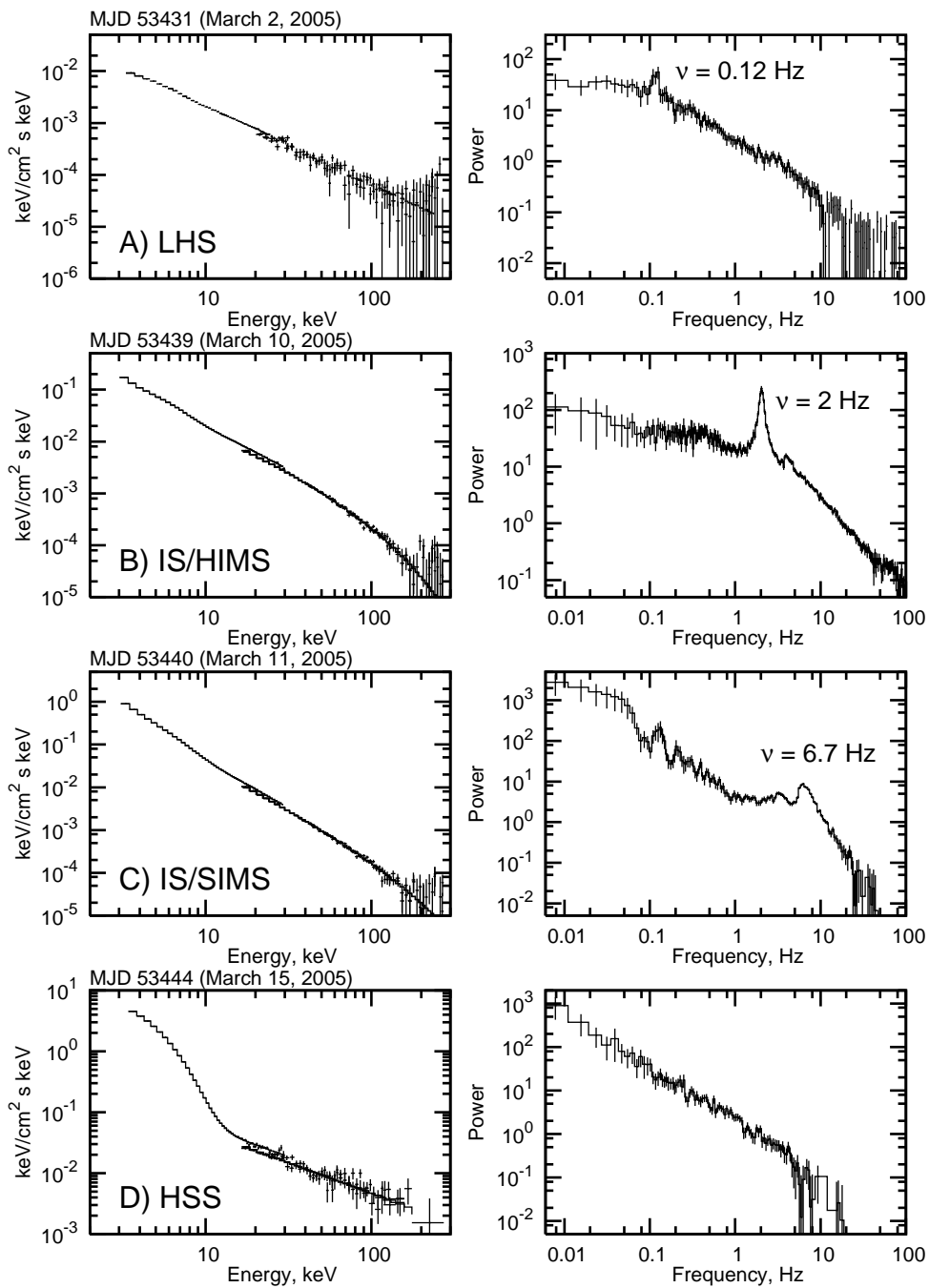}
\caption{Representative spectra of the each state observed during
the reported rise of the 2005 outburst (from the top to the bottom):
 LHS, HIMS, SIMS  and HSS.}
\label{sp_pds}
\end{figure}

The high energy cut-off in the X-ray spectrum during the IS becomes much more
 pronounced, before disappearing in the HSS. The same cut-off phenomenology 
(i.e. marginal in the LHS, strong in the IS, absent in the HSS) 
was also observed in 4U~1543--47 \citep{kal05}
and XTE~J1550--564 \citep{tom01}. The timing properties rapidly evolve
during this stage. The QPO frequency rises from 0.5 to
2 Hz. However, the overall shape of the PDS remains similar to that of the LHS, with 
increasing characteristic frequencies (Figure \ref{sp_pds}, panel B). 

\subsubsubsection{Soft Intermediate State}

\begin{figure}[ptbptbptb]
\includegraphics[scale=0.6,angle=-90]{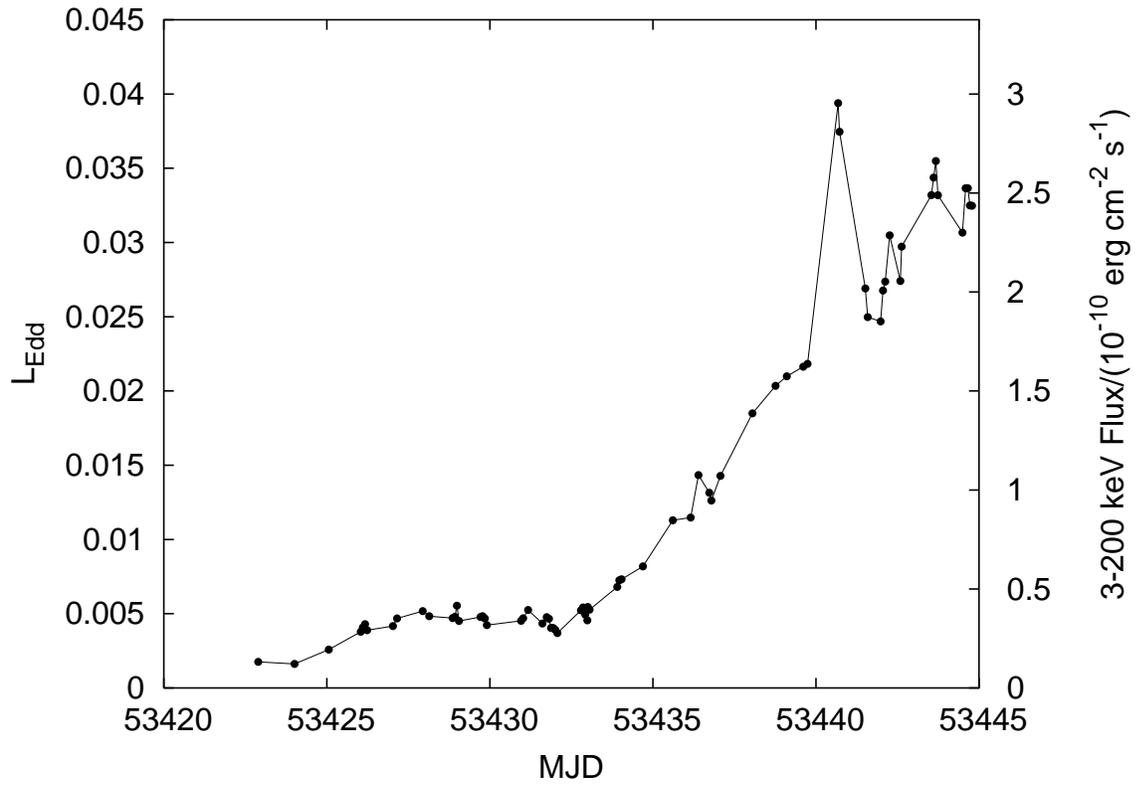}
\caption{
Source  luminosity (upper panel) in Eddington  (left scale) and flux (right scale) units.
The Eddington luminosity is calculated for the BH mass of 6.3 $M_\odot$ and accreting material 
of cosmic abundances. 
   }
\label{ledd}
\end{figure}

A remarkable quick change in the proprieties of the source emission occurred between MJD 53439 and MJD 53440 in
all wavelengths. First, the radio emission disappears and the optical flux levels off.
A sharp rise in X-ray luminosity, by a factor of 2 from 0.02 to 0.04$L_{Edd}$, occurs (see also Figure \ref{ledd}),
which is in agreement with the luminosity
for a rise-phase hard-to-soft state transition in BH sources given by \citet{mh05}.
The spectrum abruptly softens from photon index 1.7 to 2.1. The total variability drops below 10\%.  
A red-noise-like continuum feature appears above the broken power law (Figure \ref{sp_pds}, panel C). 
During the next {\it RXTE} observation on MJD 53441 the source flux recedes back to the level
consistent with the exponential rise of the HIMS and we have the 
last detection of the low-frequency QPO, at an increased frequency of
around 17\,Hz. After that the flux starts to  level off and the source
enters the HSS. 

\subsubsection{High-Soft State (MJD 53442 and beyond)}

The final stage in this early evolution is the HSS,
from MJD\,$\sim53442$ through the end of the period considered here.
This is characterized by the
rapid and somewhat jittery softening of the X-ray spectrum, the
disappearance of the radio emission and a relatively stable optical flux.
The energy spectrum is dominated by a thermal disk component and no
low-frequency QPO or band-limited noise component is detected (Figure \ref{sp_pds}, bottom panel).
 
\citet{broc06} identified MJD 53440 (March 11) as an approximate
date of the LHS-HSS transition. This conclusion is based on the behavior of the 
{\it SWIFT}/BAT light curve. Our data sampling and broad band energy coverage resolve 
distinct  HIMS and SIMS states between the LHS and the HSS. The HIMS-SIMS transition transition occurs
on 53440; our analysis is thus consistent with the \citet{broc06} estimate.

\subsection{Spectral Modeling of the X-ray data}

\subsubsection{LHS broadband X-ray spectrum}

\begin{figure}[ptbptbptb]
\includegraphics[scale=0.66,angle=-90]{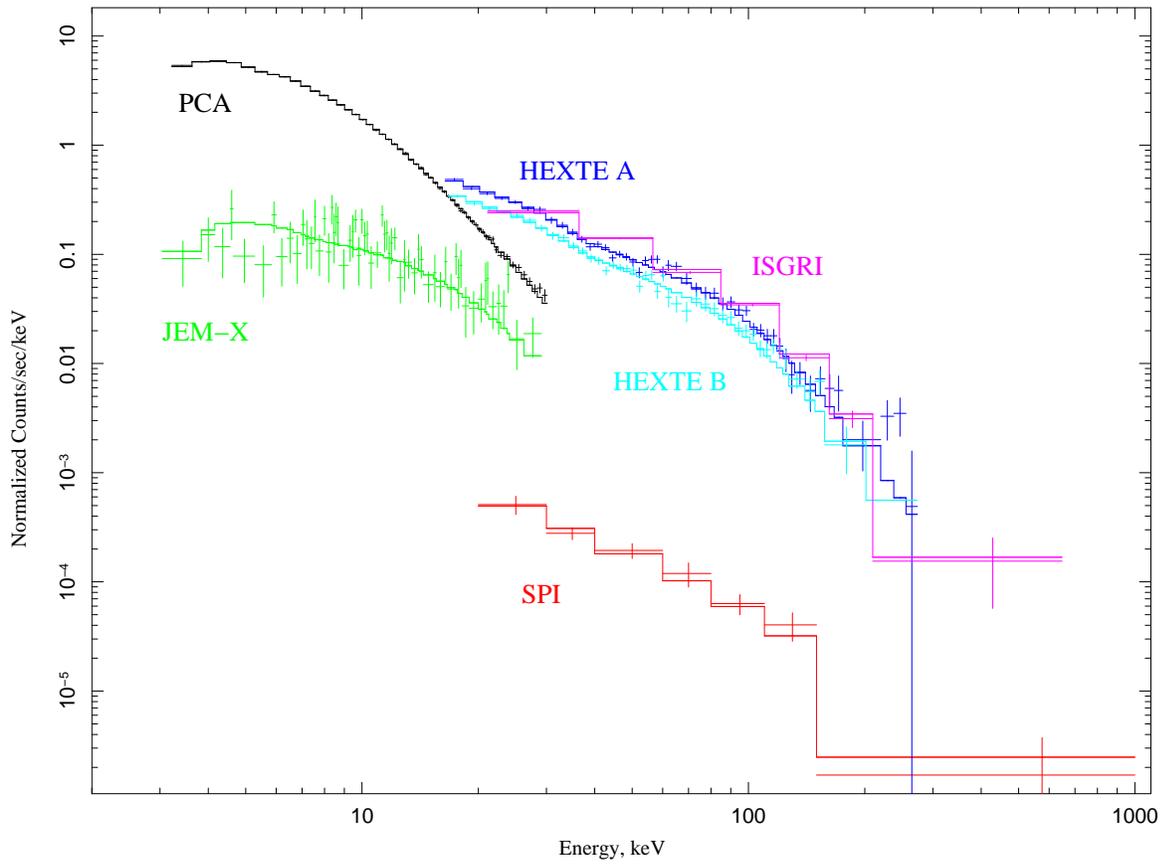}
\caption{
The broadband X-ray spectrum of the early-phase outburst by the PCA, 
HEXTE, IBIS and SPI instruments. 
The fit shown is for a black body plus a power law times a cutoff. 
 }
\label{all}
\end{figure}

During the 2.4 days of our {\it INTEGRAL} observation, the spectrum did not change noticeably. 
There were {\it RXTE} observations at the beginning, midpoint, and end of this period. We 
have used the sum of these observations (RXTE observation IDs 90058-16-04-00, 90428-01-01-00,
90058-16-05-00, and 90428-01-01-01) to construct the composite spectrum shown on Fig. \ref{all}.
The spectrum is approximately an absorbed power law, with photon index $\approx$ 1.4, 
typical of LHS spectra of accreting BHs ($\chi_r^2 = 1.86$ for 344 degrees of freedom). 
But the fit is significantly improved with more complex models. The results 
for the continuum are presented in  Table \ref{model_fits}.
We applied several different models that are often used to capture the properties
of a thermal disk and a Comptonizing corona. Here we summarize the results for 
the composite spectrum, but we discuss the models further in the following section, 
in the context of the evolution of the spectrum as the outburst develops. 
In \S\ref{discussion} we discuss the alternative of synchrotron and synchrotron self-Compton
mechanisms for the emission in the LHS. 

The fit is greatly improved with addition of a black body component,
an emission line due to iron and a high energy exponential cutoff 
($\chi_r^2 = 1.16$ for 338 degrees of freedom).
A narrow line at $6.33_{-0.13}^{+0.16}$ has equivalent width of 97 eV. 
Comptonization and reflection models both give fits of similar quality and 
the $\chi^2$ value is not grounds for favoring one over the other. 
The XSPEC BMC model, named ``Bulk Motion Comptonization'' for its applicability 
to the case of Comptonization of seed photons in plasma with high velocity bulk motion, 
employs a self consistent convolution of a Planck function with a Green's function  accounting for
Compton scattering. This should best represent the physical situation
when soft disk photons are up-scattered in a hot corona. We also 
fit the data with the model representing Comptonization in a thermal cloud 
(XSPEC's COMPTT model). In this model the effective electron temperature 
$kT_e$ is the fit parameter.  The best-fit  value of 37 keV is less than what one would expect
from the cut-off, i.e. $kT_e\approx E_{fold}/2$ [this approximate relation follows from the 
comparison of the analytical model for unsaturated Comptonization \citep{st80}
with the results of Monte-Carlo simulations \citep{t94}]. 
However,  the difference is qualitatively understandable considering the error bars on $E_{fold}$ 
and the effects on the COMPTT spectrum of the optical depth.
The model of reflection of a power law source from a cool disk (XSPEC's PEXRAV 
implemented by Magdziarz \& Zdziarski 1995)
also provides an acceptable fit to the combined spectrum.
During the fitting procedure we fixed the abundances of iron and 
heavy elements to cosmic values. We also used the inclination angle 
value $70^\circ$ determined by \citet{OB97}.
We assumed interstellar Galactic absorption column of $N_H=0.89\times10^{22}$ 
cm$^{-2}$ \citep{nh}. In modeling the combined LHS spectrum we allowed cross-normalization
constants for each instrument with respect to 
the {\it RXTE} PCA. The best fit cross-normalization values are given in Table \ref{cross-norm}.
%The normalization of the SPI detector agreed with the PCA and ISGRI , while the two other
%instruments from {\it INTEGRAL} differ by more than 20\%. The normalization
%of the HEXTE clusters agreed with the PCA within 10\%. 

\subsubsection{{\it RXTE}/PCA spectral evolution}

In the disk plus corona model, thermal 
radiation is presumably generated in the inner accretion disk. Some portion 
of these photons are Compton scattered in a hot, optically thin media (corona) 
situated above the disk.  
The photons escape diffusively, with electron 
scattering being the dominant source of opacity.  The spectrum observed at 
infinity consists of a soft component coming from input photons that escaped 
after a few scatterings and of a power law that extends to high energies 
comprised of photons that underwent significant up-scattering. 
%Iron fluorescence line and absorption edge features 
%may be observed in the energy range 5-9 keV. 
The spectrum emitted by a part of the accretion disk not covered by a corona is thermal 
and ideally should be represented by either a black body shape  with 
$kT_{col}\approx0.5 - 1.0$ keV or an integral over temperatures for the  
multicolor disk model (MCD, DISKBB in XSPEC), \citep{diskbb}.  
However, we found that real situation is more complicated especially in the
HSS where we need two thermal components to fit the spectrum
(see discussion below).

%Photons from the inner accretion disk are partly 
%Comptonized by the covering corona and form a power law spectrum.  

The {\it BMC} model 
in XSPEC was developed as a generic Comptonization model by \citep{tmk97}
to treat Compton upscattering of low-frequency photons in a converging flow 
of thermal plasma onto a black hole. It can describe the production of a steep 
power law component in the BHC HSS. 
Soft seed photons in the BMC model are assumed to have a pure black body energy distribution with
temperature $kT$. A fraction $A/(1+A)$ of the input spectrum is Comptonized 
in the hot corona, while the other part leaves the system unchanged. However, in 
the lower $\dot{M}$, low-hard state configuration, reduced Compton cooling of 
ambient plasma leads to electron temperatures much higher than that of the 
converging inflow. The thermal source is then not seen, and scattering from
the higher temperature plasma dominates the observed emission.  The {\it BMC} 
model is applicable in the general case of Comptonization when the 
bulk and thermal electron motion are included. This model can be used 
provided the electron temperatures involved are higher than the mean photon energy. 
At higher photon energies,  the spectrum
exhibits a cutoff at 100-200 keV, indicating the photon energies are on the order 
of the electrons thermal energy, and the cutoff multiplying the {\it BMC} model represents 
the effect the electron energy distribution would have.  
This model is convenient for describing the evolution of a BHC through different states 
and is based on a theoretical picture. 

The reflection model {\it PEXRAV} along with {\it DISKBB} can also describe the spectral changes 
of BHCs, although it does not itself provide an underlying rationale for the changes 
that occur. In principle, effects of reflection should be added to the {\it BMC} model. 
However, the models are too much alike in functional form 
for the fits to the data to separate the contributions to the continuum. 
 
To examine the time evolution of the energy spectrum, we analyzed 
{\it RXTE}/PCA data from the beginning of the outburst until  March 15, 2005
(MJD 53444), when the source has completed the LHS-to-HSS transition.
Following the above discussion, we fit the data to  {\it BMC+GAUSSIAN} and {\it PEXRAV+DISKBB+GAUSSIAN} models.

\begin{figure}[ptbptbptb]
\includegraphics[scale=0.8,angle=0]{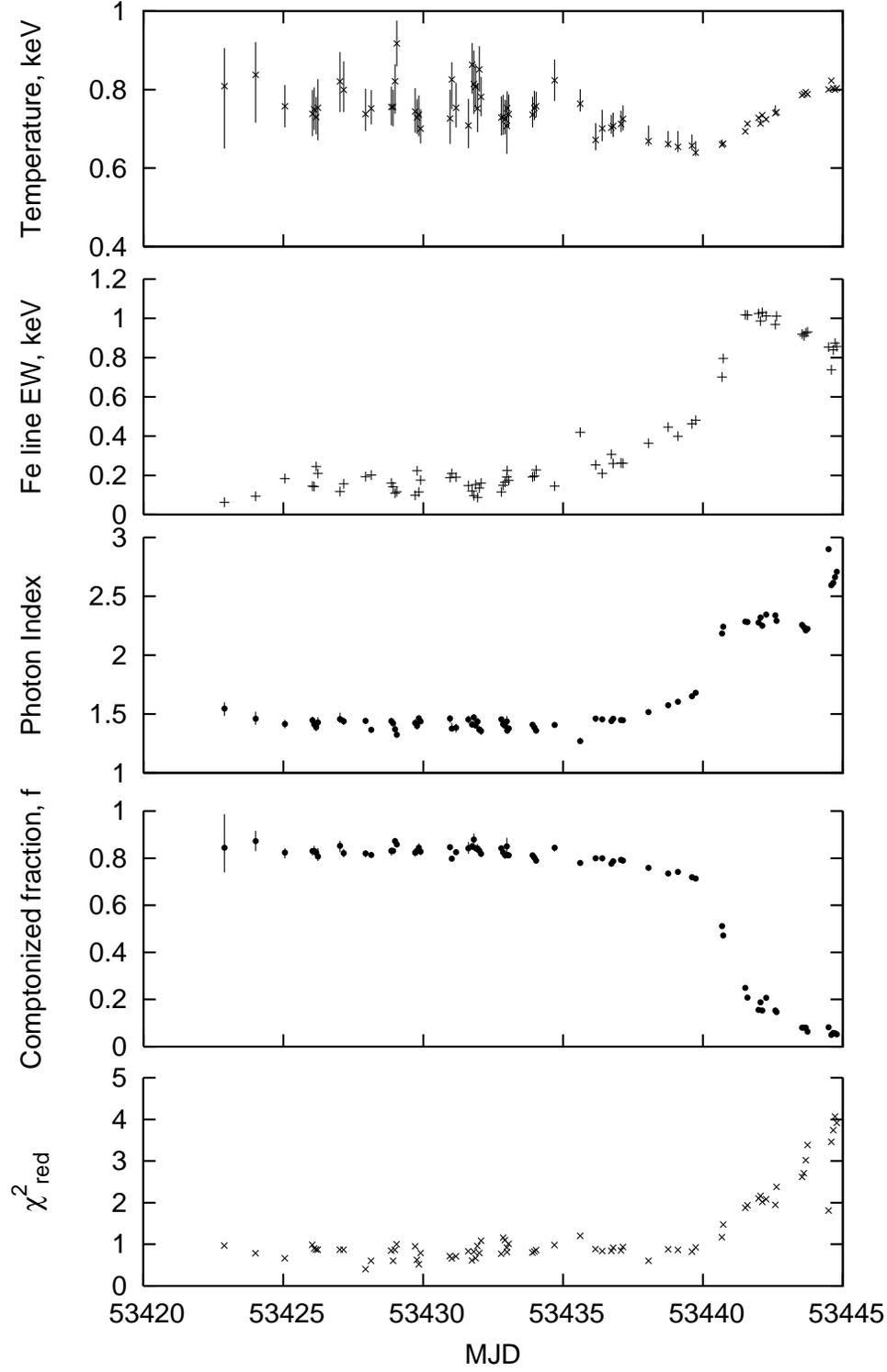}
\caption{Outburst spectral evolution: {\it BMC} model component parameters.
The model describes the data well in LHS and HIMS, while 
in the HSS the fit is not satisfactory. }
\label{bmc}
\end{figure}

\begin{figure}[ptbptbptb]
\includegraphics[scale=0.7,angle=0]{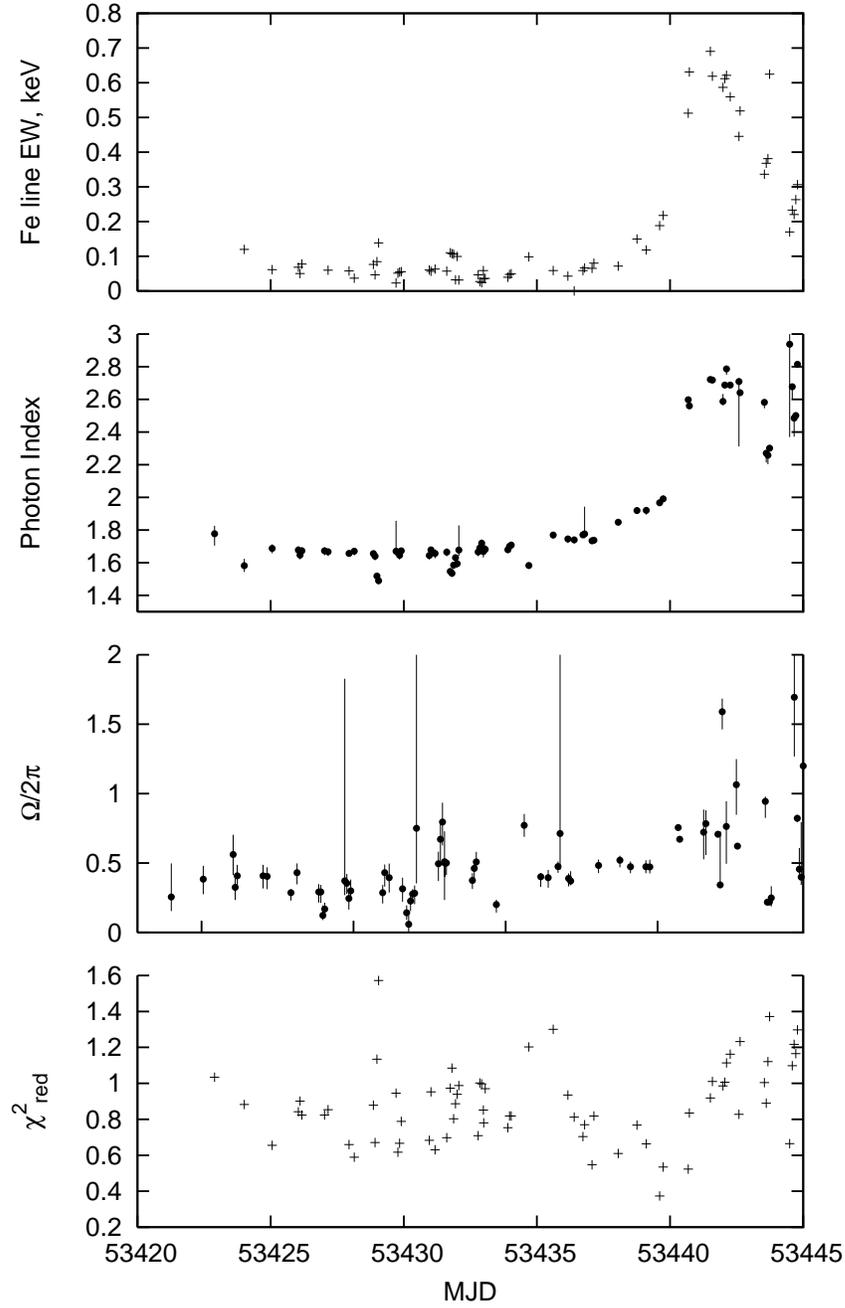}
\caption{
Outburst spectral evolution: {\it PEXRAV}\  model component parameters.
In the text the difficulty of explaining the iron line is discussed.}
\label{pexrav}
\end{figure}

The results are given on Figures \ref{bmc} and \ref{pexrav}.
 for {\it BMC} and {\it PEXRAV} model application correspondingly. 
In fitting the {\it PEXRAV} model 
we let the photon index and reflection factor 
change freely and fixed the abundance of iron and heavy elements at 1.0. 
The spectral index given by fits to the {\it BMC} model rises from the value 
of 1.4 for the LHS to  1.8 during 
the state transition, and reaches 2.2 for the HSS, while the {\it PEXRAV} model results in 
power law spectral indices 0.2-0.4 higher.
The disk component is insignificant during the LHS and the {\it entire} IS for the {\it PEXRAV} model.
For both models a Gaussian with
the centroid energy of $\sim6.5$ keV was included to model an iron line. 
For the LHS the iron line is mostly narrow (sigma $\leq$ 1 keV). The  equivalent width (EW)
of the line is $\leq$ 200 eV for {\it BMC} and $\leq 100$ keV for {\it PEXRAV},
both  within the range expected from the cold disk reflection. However,
after the transition, the Gaussian derived in the fits becomes much broader, with sigma 1.5--2.0 keV and
the EW of the feature grows by a factor of 3-5. 
Reflected iron lines have been calculated for a variety of conditions. 
Equivalent widths this high, and also very broadened lines features, can be obtained
\citep[e.g.][]{dab01,mar00,bal02}, but only for special extreme 
conditions of position of the source relative to a maximally 
rotating black hole or high over-abundance of iron. 
High iron line EWs were obtained in models of  reprocessing in a wind
from the central source \citep{lt04}, but this model did not give large widths 
and is besides not thought to be applicable to the HSS.
Alternatively, the high width of the Gaussian may indicate that
the model component, initially intended to model a narrow spectral line,
is trying instead to mimic some part of the broader underlying continuum
not included in the model. Our analysis also shows that a pure BMC model
(with additional {\it GAUSSIAN} fixed at 6.5 keV for an iron line) during 
the HSS is not statistically acceptable ($\chi^2/N_{dof}$\gax 2.0).
We attribute this behavior to the fact that the accretion disk emission is
not well described by a pure black body. We attempted to resolve the issue by
replacing the unscattered black body component of the {\it BMC} component by a multi-color
disk component (i.e. {\it DISKBB}). The quality of fit was slightly  improved;  however the
behavior of the Gaussian remained the same, with very high widths and EWs required.

\begin{figure}[ptbptbptb]
\includegraphics[scale=0.7,angle=0]{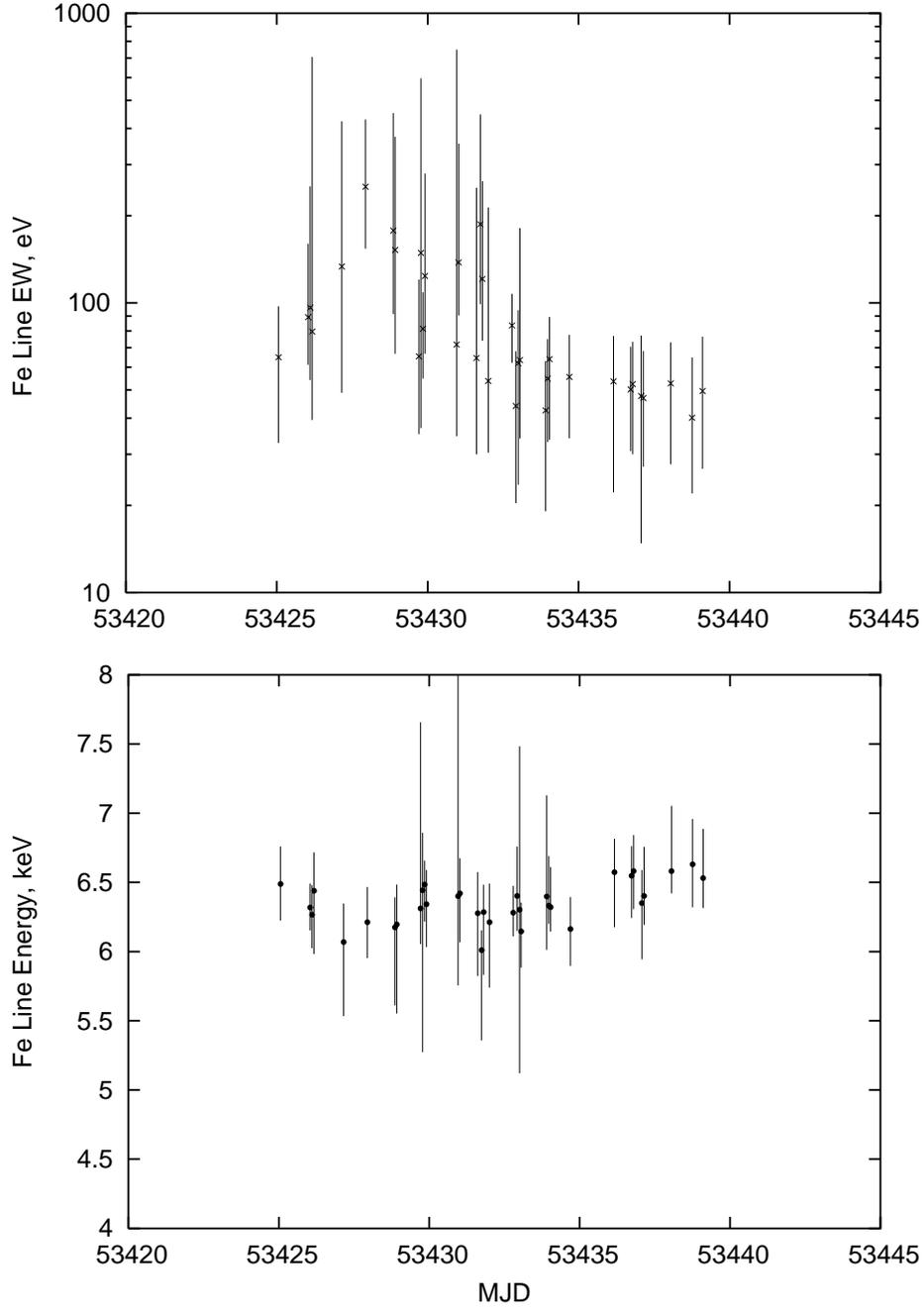}
\caption{
The iron line properties inferred from the  {\it BMC + BLACK BODY} fits.
The line is narrow. The 3$\sigma$ upper limit on the 
flux in the HSS corresponds to an upper limit on the equivalent width of only 9 eV. 
 }
\label{bmcbbln}
\end{figure}

Secondly, we replaced the Gaussian in the {\it BMC+GAUSSIAN} model by a simple black body shape. 
The model fit was dramatically improved. We obtained the following properties 
of this alternative model. 
The temperature of the new soft thermal component
is approximately half of the {\it BMC}'s seed photon temperature.
 The temperature of the seed photons, in turn, slightly increased,
compensating for the line-like continuum residual.
 The narrow iron line is only needed for 
the LHS and the IS (see Figure \ref{bmcbbln}).
 For the HSS, a good fit is achieved  with {\it BMC} and soft black body only. 
 Interestingly the narrow iron emission line, 
with an energy consistent with fluorescence increases in flux along with 
the increase in the non-thermal spectrum that might excite it from the
disk material and disappears when the spectrum becomes steep and dominated by the 
thermal emission. A soft thermal component might  be  
emission from larger radii of the accretion disk than the source of seed photons
for the BMC. Its temperature $kT_{bb}$ is consistently lower than the temperature 
of incoming photons for the BMC component.  In fact, the relation $kT_{bb} 
\sim 0.5 kT_{BMC}$ holds throughout the entire time period of interest.
The  soft black body component is 
barely significant for the LHS and its contribution 
grows towards the HSS. This fact is consistent with the conventional 
scenario, in which  the size of the corona shrinks as the accretion rate  
increases and the source enters the state transition. The size of the 
uncovered disk increases, allowing for a greater contribution of the soft
component. The temperature of this component grows with approach to the HSS,
which is also in agreement with the implied picture as inner 
parts of the disk should supply the hotter photons.
From the same reasoning it is clear why the multi-color disk does not 
work in this case. In the {\it DISKBB} model, the
normalizations of the inner and outer parts of the disk are
locked together by the integration over the radius
with the $r^{-3/4}$ temperature profile. In the real situation, a part of the radiation
from the inner component is taken away by Comptonization. This imbalance is not
accounted for the XSPEC {\it DISKBB} model.

Recently, several models were developed which take into account effects not handled correctly
in the {\it DISKBB} model. The {\it KERRBB} model \citep{kerrbb} is the most inclusive.
We tested this model
on the  observation during the HSS  on MJD 53443 (March 14) 
(RXTE Observation  ID: 91702-01-03-00).
We fit the PCA spectrum with a model consisting of {\it KERRBB+BMC} with 
the thermal component of the {\it BMC} model
suppressed (by fixing log(A) parameter at 7.0). For the {\it KERRBB} component 
we fixed the inclination angle 
at 70$^\circ$, the distance at $3.2$ kpc and the BH mass at 6.3 $M_{Sun}$ and f$_{col}$=1.65. 
We also fixed the parameter 
$eta$ at zero, which corresponds to a standard Keplerian disk with zero torque at the inner boundary.
The best fit gives $\chi^2_{red}=2.18$. Addition of a Gaussian with the energy fixed at 6.5 keV
results in an acceptable  $\chi^2_{red}=0.98$. The equivalent width of the feature is a reasonable 210 eV.
The best fit value of the
specific angular momentum of the black hole is $0.53\pm 0.01$, somewhat less than the 0.65-0.75 
found by \citet{shafee06} with the same f$_{col}$ for {\it RXTE}'s 1997 observations. 
More detailed modeling would be needed to determine whether the line width of 1.2 keV 
could be accommodated in terms of Doppler shifts and reddening. 
{\it KERRBB} can handle 
corrections to the basic disk picture that are known to be needed and these seem to be on the
order of the discrepancy between the data and the simple disk approximations. It 
needs to be used in attempting to separate line emission, reflection, and possible structure of 
disks, as well as to determine the black hole's angular momentum.

\subsection{Intensity Dips}

In the 1996/97 outbursts, when the source was in the HSS, a variety
of dips were observed (Kuulkers et al. 2000 (K00 hereafter); 
Kuulkers et al. 1998;
also see \citet{tan03}, regarding dips during the 1994 outburst). In the
case of our current campaign, the source remained for the most part in
the LHS. Nonetheless, examination of the 16-s standard mode
light-curve data also revealed dips. This is a little
surprising, as the dips are widely believed to be associated with partial
occultation of the central disk region by the accretion stream or by more complex geometrical
accretion configurations. The LHS emission on the other hand, consists of
Comptonized photons, presumably scattered from plasma ambient to the disk.
There have been arguments that this scattering media is large compared to the
inner disk region where the thermal X-radiation emanates \citep[e.g.][]{hua99}
That would seem to make the mechanism which produces the HSS
dips less effective for the LHS case. In any case, K00
find that for the case of the 1996/97 outburst, the HSS dips occupy a
relatively narrow, approximately Gaussian distribution when plotted as a
function of phase. Specifically, they find that the dips are 
approximately described by a Gaussian centered on phase 0.785  with a sigma of 0.046.

\begin{figure}[ptbptbptb]
\includegraphics[scale=0.6,angle=-90]{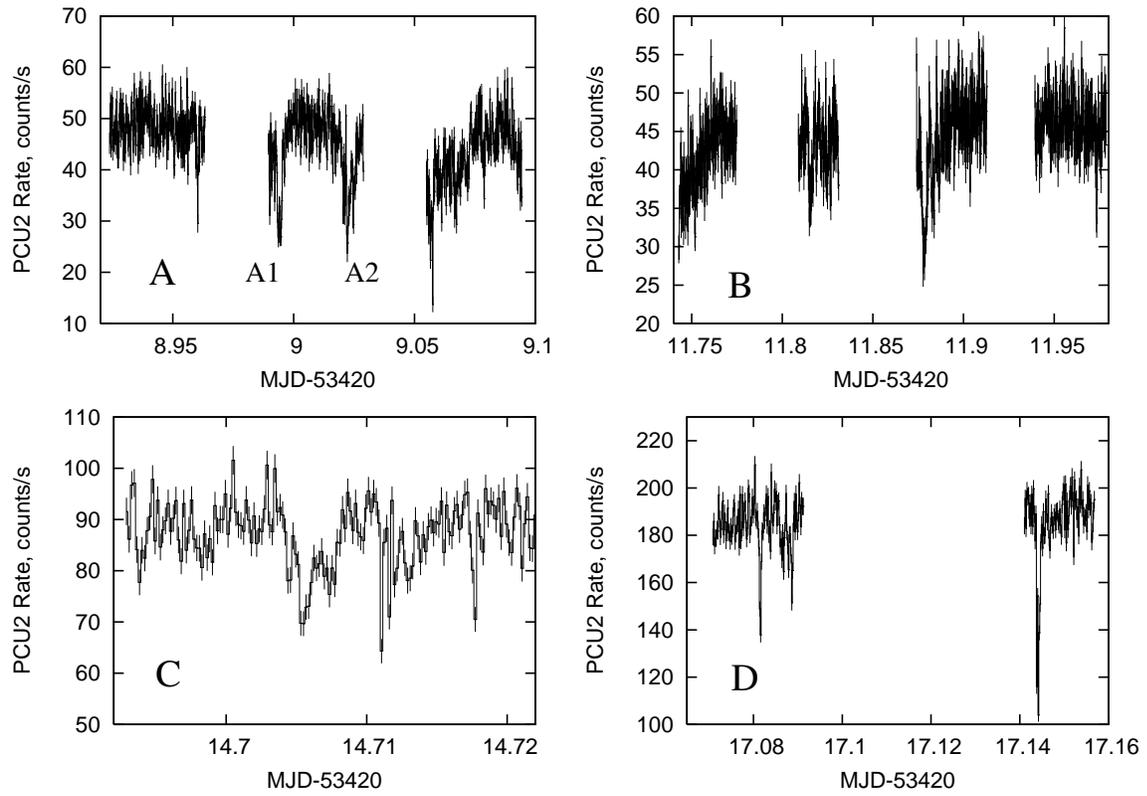}
\caption{
Examples of dips observed in PCA light-curve. The average total rate (2--60 keV) per PCU 
is shown, with 16 s time resolution.
 }
\label{dips}
\end{figure}

To study the dipping behavior in our data, we examined the entire 16-s 
standard-mode light curve, which of course has gaps;
see Figure \ref{dips} for examples. For the ephemeris calculation we used the 
zero spectroscopic phase 
2,449,839.0763$\pm$0.0055 JD of \citet{OB97} and the 2.62191$\pm$0.00020
day binary period of  \citet{greene01}, which is calculated consistently 
with the \citet{OB97} data. We shifted by 0.75 phase to obtain an equivalent
photometric phase zero of 2,449,841.0427$\pm$0.0055.
The $2\times 10^{-4}$ day uncertainty in the period, combined in 
quadrature with the uncertainty in the phase zero determination translates
to an uncertainty of 0.11 phase in our data. 

\begin{figure}[ptbptbptb]
\includegraphics[scale=0.6,angle=-90]{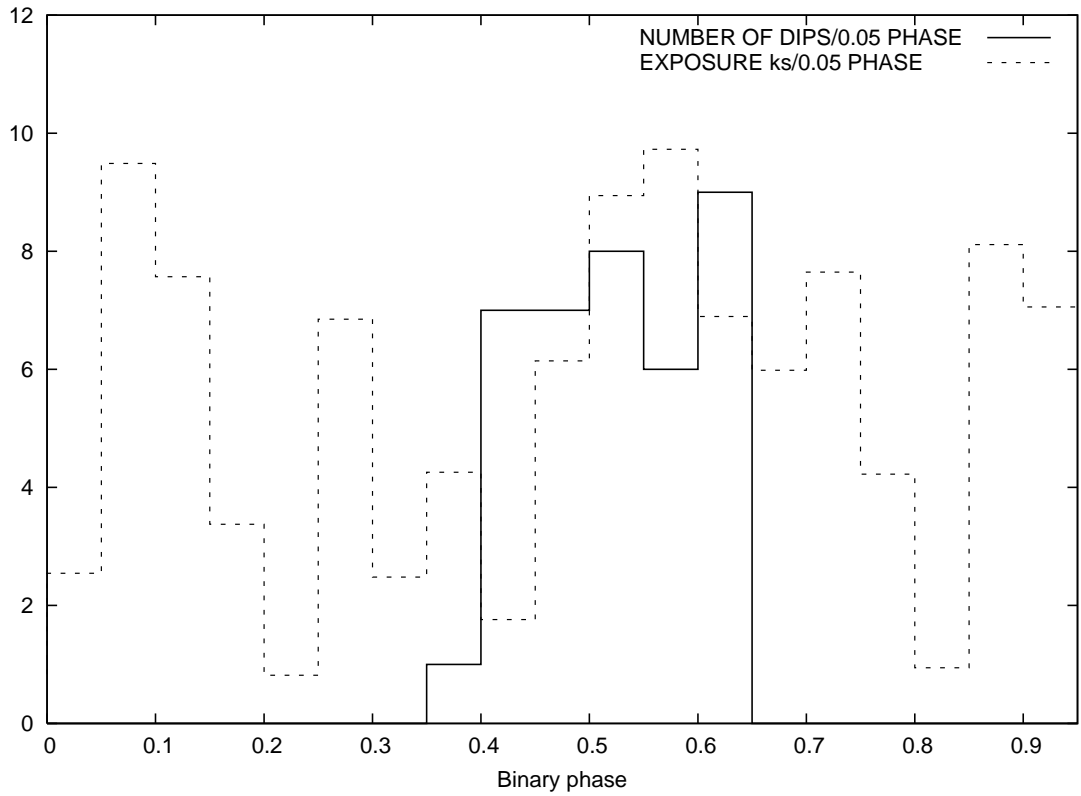}
\caption{
Distribution of light-curve dips as a function of the binary orbit phase (solid line).
Cumulative exposure per phase bin is shown by a dashed line.  
Previous studies, conducted while the system was in the HSS, found a dip-phase  distribution
that was approximately Gaussian, centered at 0.785 \citep{erik00}.}
\label{dip_hist}
\end{figure}

We identified
instances where the intensity dropped by $\sim$50\% or more, relative to the
local average, in at least one bin, at a significance level of at least
5 sigma. We found  46 dips in total, 28 in the LHS, 10 in the IS, 
and 8 in the  HSS. We then sorted the dips into orbital-phase bins using the 
ephemeris information as described. This resulted in the 
distribution shown in Figure \ref{dip_hist}.
Note that while our dips are still relatively narrowly distributed in
phase, they are centered on phase 0.5, with substantially more dispersion than 
K00 found for the 1996/97 case. But even with the 0.11 phase uncertainty the 
distribution appears shifted significantly (at about the 2$\sigma$ level) 
in phase relative to K00. Furthermore, we find no dips within the 
narrow phase-0.79 centered distribution of K00 at a $\sim$1.5-$\sigma$
level of confidence.

\begin{figure}[ptbptbptb]
\includegraphics[scale=0.65,angle=-90]{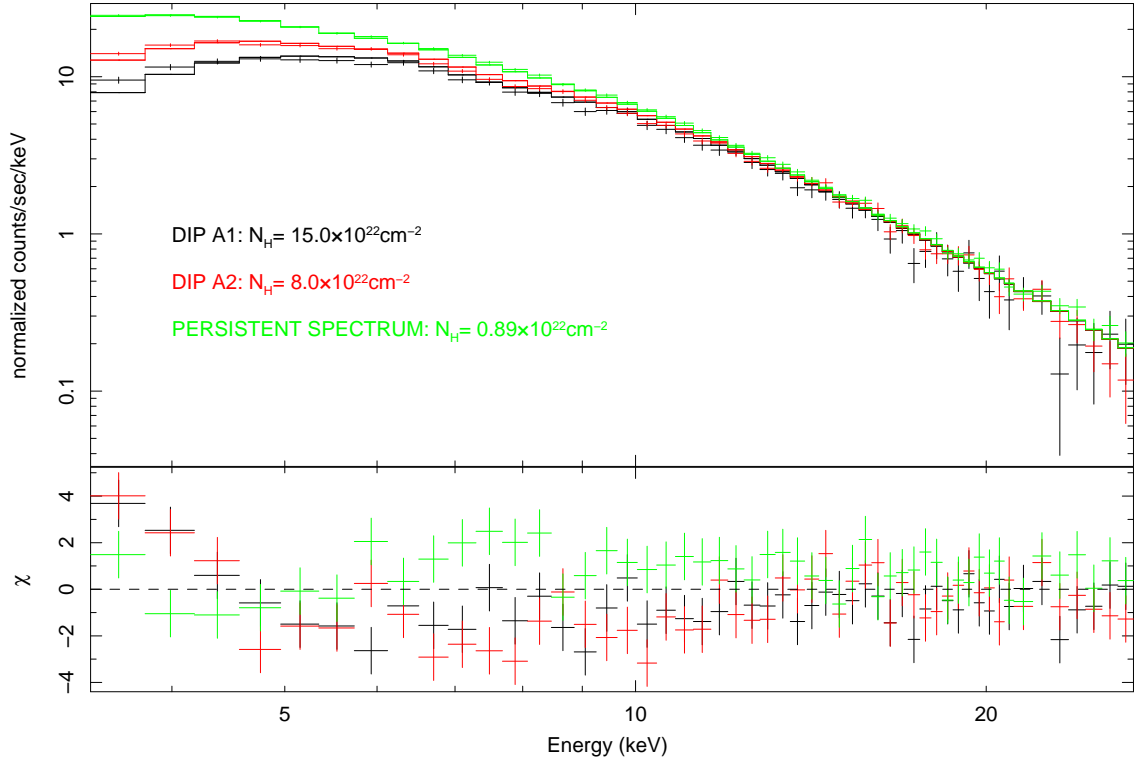}
\caption{
Absorption nature of the dips observed in hard state.
Spectra shown in black and red are extracted during dips A1 and A2 (see panel A on Figure \ref{dips}) while spectrum for the 
interval between the dips is plotted in green. The change in spectrum
during dips is accounted for by variation in absorption column $N_H$ . }
\label{dip_spec}
\end{figure}

We do not
have as large a statistical sample as K00 (46 versus 65). Nonetheless, it
is interesting that K00 observed no intensity dips near phase 0.5-0.6, or
any other phase significantly different from 0.8. This is at least
suggestive of a different mechanism in the LHS configuration of the system. Another
difference is that our spectral analysis finds that pure
photoelectric absorption provides an adequate explanation (see Figure \ref{dip_spec}). 
In addition, we examined public data spanning approximately from 
MJD 53445--53460, i.e. after the onset of the high-soft-state transition. 
We find a dearth of dipping activity in that portion of the data. 

We note that Tanaka et al. (2003) reported unusual dipping behavior during 
an ASCA observation of GRO~J1655--40 in 1994. In that case, the dip profile was significantly
broader, with a width of several hours, while ours are typically
several 16~s bins wide. They reported that it occurred in the HSS when the 
secondary was on the far side of the black hole, and that their spectral 
analysis was consistent with a multi-layer partial covering. Most
of the dips we observed are not in the HSS (though a few are) and 
our spectral analysis is consistent with pure photoelectric absorption
by a single phase medium. Thus, we tentatively conclude that we are seeing
events of a distinct nature from that observed in 1994. Speculatively, it 
could be that we are sampling the effects of narrowly collimated
accretion stream(s) while Tanaka et al. (2003) observed effects associated 
with irregularities in the geometrical configuration of the outer disk. 

\section{Discussion and Conclusions\label{discussion}}

We have studied  multi-frequency data spanning the early stage of the 2005 
outburst of GRO J1655-40. The  empirical behavior observed includes:
spectral transition from the LHS through the HIMS and SIMS to the  HSS, accompanied by
the expected rise in the thermal component and a steepening of power law spectral
index; QPO appearance and its frequency rise throughout the LHS and the intermediate states and its
turnoff upon entering the HSS; a  high-energy cut off in the LHS and IS;
intensity dips during the LHS 
which are consistent with photo-electric absorption and correlated differently with orbital phase
than the  HSS dips observed by K00 for the 1996--1997 outburst. 

We observe rising radio emission  during
the LHS, a declining radio flux in the HIMS and its turn off in the SIMS.
The radio data from MJD\,53433 are unique in showing a
pronounced peak at 5\,GHz.  X-ray binaries in the LHS typically
produce a very stable, flat-spectrum radio source, as seen here in the
plateau stage.  A 5\,GHz peak is highly unusual, and suggests an expanding,
optically-thick source, typical of the early stages of jet ejection.
The radio coverage allows only poor limits on the ejection
date (later than MJD\,53429, earlier than 53433), but it is tempting to
associate this ejection with the abrupt change from a steady 
to a sharply brightening X-ray corona.  In this context it is also
interesting that during the X-ray exponential rise,
the radio flux was comparatively steady, with a marginal steepening
of the  spectrum.  While the latter is not convincing on any given day,
the radio spectrum is
consistently flatter (harder) during the plateau stage, and steeper (softer)
during the exponential rise.  Taken together with the decoupling between the
X-ray and the radio {\it flux}, it seems likely that the radio emission here
has a different physical origin from that during the plateau.  The plateau
seems much like the standard LHS, in the ratio of X-ray to radio
fluxes, and the flat radio spectrum, suggestive of a compact core. The
radio emission during the HIMS may instead be a more extended
jet, though it is far from obvious why that jet should be so stable (in flux
density and spectrum). 

\begin{figure}[ptbptbptb]
\includegraphics[scale=0.6,angle=-90]{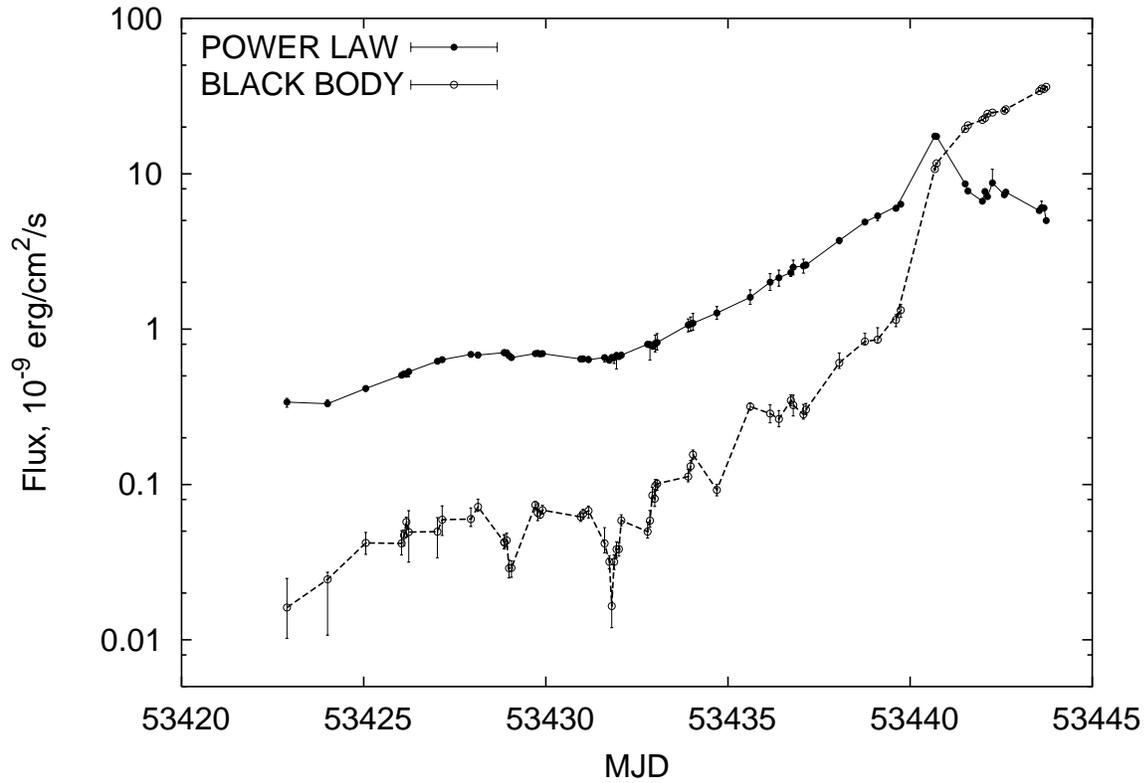}
\caption{ Energy fluxes in spectral components for fits to the 
{\it POWERLAW + BB} model. The power law flux is calculated 2--20 keV.
The black body  flux is the bolometric flux inferred from the normalization 
of the component. An apparent jump occurs in both fluxes around MJD 53441. 
After that the black body component continues to rise, while the power law flux 
starts to decline.
}
\label{flux_vs_time}
\end{figure}

Currently, there are two competing theories explaining the
origin of the X-ray power law spectrum in BH sources. In the first, the ``canonical'' 
model, the power law slope is produced by Comptonization of the soft disk
photons in the extended optically thin corona, located above the accretion
disk. But \citet[][and references therein]{mar03},
argue that the power law component in the LHS can be attributed 
to a combination of 
the optically thin high energy tail of the synchrotron 
spectrum from the radio jet base and synchrotron self-Compton emission
adding the highest energy component. In the framework of the  model,  
the synchrotron jet base subsumes the role of the corona in 
Comptonization models \citep[e.g.][]{nowak05,markoff05}.
 However, the lack of radio and X-ray correlation in our
data  make it difficult in the case of GRO~J1655--40 to
associate the base of a compact jet with the Compton-scattering corona which
creates the power law X-ray emission.  
\citet{gallo} found evidence for radio and X-ray correlation for several 
sources, though they also saw that in some cases the radio was quenched at 
some point in the X-ray rise.
It is possible that the radio emitting region is larger than the X-ray corona,
and not dominated by the X-ray emitting part. 
This would mean that the situation may be more complex.
Recently the jet-base model has been subject to other observational challenges.
\citet{mac06} argued that if the LHS spectrum is due to the jet, which is 
radiatively inefficient in nature, the transition to the HSS should 
be accompanied by a sharp change in luminosity. While we observe a rather
quick rise in the thermal component during the SIMS around MJD 53440 
(see Figure \ref{flux_vs_time}), the corresponding power law flux does not drop, but
 jumps up together with black body component before it starts to decay smoothly.
 Overall, our analysis does not validate the jet-base
model; however some presence of synchrotron and synchrotron self-Compton emission can not be ruled out.

\begin{figure}[ptbptbptb]
\includegraphics[scale=0.8,angle=-90]{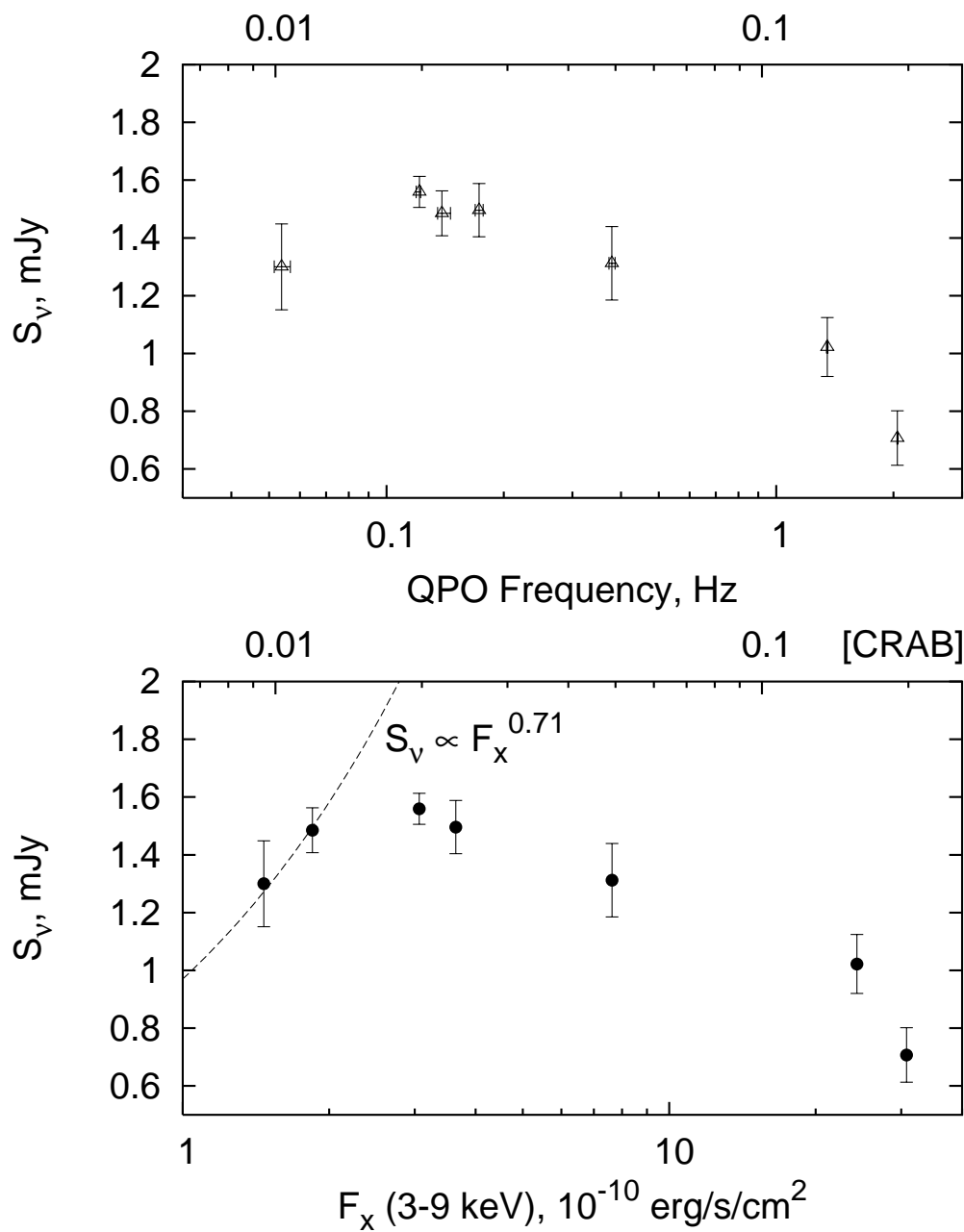}
\caption{
Upper panel: 8.46 GHz radio flux dependence on QPO frequency.
The data do not show the linear correlation derived by \citet{mig05} for
GX 339-4.  
Lower panel: Radio flux versus 3-9 keV X-ray flux. 
The data clearly departs from the
$S_\nu\propto F_x^{0.71}$ dependence obtained by \citet{gallo}.
}
\label{flux_corr}
\end{figure}

\begin{figure}[htbp]
\includegraphics[scale=1.0,angle=-90]{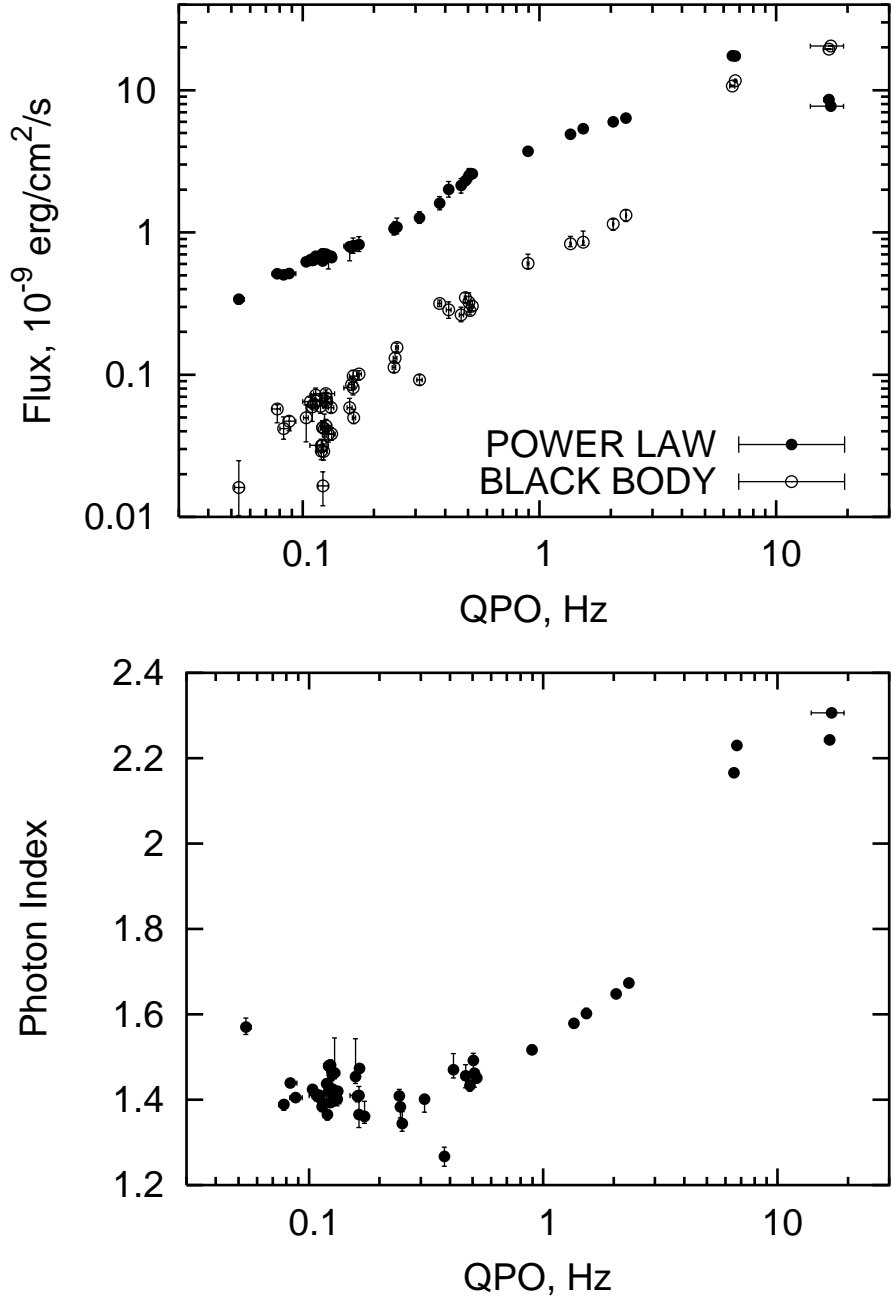}
\caption{Upper panel: observed correlation between the QPO frequency and fluxes 
in the {\it POWERLAW + BB} model. The fluxes are calculated as in Figure \ref{flux_vs_time}.
Lower panel: the behavior of the photon spectral index as the QPO frequency changes.}
\label{qpo_alpha}
\end{figure}

It is apparent that the system is far from a steady state configuration 
during the LHS and the state transition.  Most probably, the accretion disk, which is quiescent prior
to the outburst, gradually gets involved in outburst activity by some instability
propagating either  from the innermost parts of the disk to the outside
or from the outer disk to the inside\citep[e.g.][]{can95}.
Information from the QPO phenomena can give an important clue
towards distinguishing between the two possible directions of the heat 
front propagation. For example, in the case of shock-wave or front propagation in the
disk one could expect the X-ray flux to be modulated at the frequency of
the disk Keplerian rotation at the radius of the shock location. If this process
is indeed responsible for the observed flux oscillations, then the outward
propagation scenario is not correct and the shock is propagating inward from larger radius,
as advocated by \citet{ch05}. 
However, there is evidence for a closer  coupling of the QPO phenomena with the power law component
than  with the disk component, so that the accretion disk shock wave oscillation model
for the oscillations is not compelling. QPO rms variability is found to
to increase with energy \citep[see][and references therein]{klis95}.
\citet{vig03} found a close correlation of power law spectral index
 with QPO centroid frequency for  BH sources GRS~1915+105, GRO~J1655--40,
 XTE~J1550--564 and 4U~1630--47. For GRO~J1655--40, Vignarca et al. used 
 data from the 1996--1997 outburst. Figure \ref{qpo_alpha} shows that in the 2005 
outburst rise the spectral index was correlated with 
QPO frequency above about 0.8 Hz. Both the power law 
and the black body (seed photon source) fluxes were correlated with the
frequency, while the radio flux is not (see Figure \ref{flux_corr}).
In the BMC model, the power law flux and the black body are related.
However, the spectral index is independent of QPO until a significant rise
in flux has occurred. Qualitatively, these correlations are consistent with 
the BMC model and do not exhibit the same indications of a jet-base model
shown in some other sources by \citet{mig05}. 
There are other models that could relate this QPO frequency to the fluxes, such as 
that of the occultation of the power law component by a precessing disk ring \citep{schn06}.
The dynamical consistency of such a ring and the connection to the spectrum of the disk
component should be further investigated. 

The location of dips has been discussed in terms of the impact of the accretion stream on the disk
\citep[e.g.][]{bis05}. The location may be a function of the rate of accretion flow. The 
structure of the dips may put limits on the size of the emission region,
as has been done for neutron star low mass binaries \citep[e.g.][]{chur01}. Detailed
study of this is beyond the scope of this paper.

In summary, the 2005 outburst of GRO~J1655--40 was observed intensely
with X-ray instruments, with very interesting and diagnostic results.
During the early phases there was good radio coverage, and less
thorough, but nevertheless useful optical coverage.  The X-ray
data allow spectral fits to generic Comptonization and power law plus reflection models
with a cutoff, multicolored disk and fluorescence. 
If the corona is really the base of a relativistic outflow, the radio flux does not exhibit
the proportionality to the X-ray flux that is seen in some other BH sources,
nor a simple anti-correlation of the flux between the power law and the disk components
of the spectrum. Low frequency QPO seem to suggest that their origin, presumably
at an interface between an optically thick and a coronal component, is
moving inwards in radius, as the frequency is positively correlated
with the luminosity in the power law, the luminosity in a disk
component (and inversely with its inner radius), while the
radio flux decreases as the QPO reaches its maximum. The spectra of the power 
law component are limited by a high energy cutoff, although 
we are unable to study this aspect in detail. While our study in this paper extends
only to when the HSS began, the observations that followed will
address many other points, including the degree of similarity of the
decay of this transient to its rise.

We wish to thank J. Homan and collaborators for the use of the data
from their proposal and  E. Kuulkers and the ESA INTEGRAL Science Operation Center
(ISOC) for providing the on-line Galactic Bulge Monitoring database.

\newpage

\newpage

\newpage

\newpage

\newpage

\newpage

\newpage

\newpage

\newpage

\newpage

\newpage

\newpage

\newpage

\newpage

\newpage

\newpage

%
%%%%%%%%%%%%%%%%%%%%%%%%%%%%%%%%%%%%%%%%%%%%%%%%%%%%%%%%%%%%%%%%%%%%%%%%%%%%%%%%
%
%\clearpage
\begin{deluxetable}{c c c c c c c }
\tablecolumns{7}
% \tabletypesize{\footnotesize}
\tabletypesize{\small} 
\tablewidth{0pt}
% \rotate
\tablecaption{Radio Observations
  \label{vladata}
  }
\tablehead{
  \colhead{    } & \colhead{   } & \colhead{$\nu$\tablenotemark{c}} &
  % \colhead{} & \colhead{Flux} & \colhead{Phase} &
  \colhead{$S_\nu$\tablenotemark{d}}  &
    \colhead{$\int S_\nu\, d\Omega$\tablenotemark{e}} &
    \colhead{$\sigma_\nu$\tablenotemark{f}} &
  \colhead{$dt$\tablenotemark{g}} 
  % & \colhead{ }
  \\
  \colhead{Date\tablenotemark{a}} & \colhead{MJD\tablenotemark{b}} &
    \colhead{[MHz]} &
  % \colhead{Telescope\tablenotemark{d}} &
  %  \colhead{Calibrator\tablenotemark{e}} & \colhead{Calibrator} &
  \colhead{[mJy/beam]}  & \colhead{[mJy]} & \colhead{[mJy/beam]} &
  \colhead{[min]}
  % & \colhead{Notes\tablenotemark{j}}
  }
%
%%%%
%
\startdata
%
%%%%
%
% \sidehead{ GRJ1655 }
  2005 Feb 23\tablenotemark{h}    & 53424.5786       &       \phn1425.0 &

               \phn\phn\phn$1.63$\phn &            \phn\phn\phn$1.73$\phn &              \phn0.11 &
       \phn21.3 \\
  2005 Feb 27\tablenotemark{i}    & 53428.5184       &       \phn1425.0 &

               \phn\phn\phn$1.90$\phn &            \phn\phn\phn$1.83$\phn &              \phn0.17 &
    \phn\phn8.9 \\
  2005 Mar \phn4 & 53433.5088       &       \phn1425.0 &

               \phn\phn\phn$1.15$\phn &            \phn\phn\phn$1.18$\phn &              \phn0.22 &
    \phn\phn4.5 \\
  2005 Mar \phn6 & 53435.5149       &       \phn1425.0 &

               \phn\phn\phn$1.78$\phn &            \phn\phn\phn$1.84$\phn &              \phn0.20 &
    \phn\phn7.7 \\
  2005 Mar \phn9 & 53438.5695       &       \phn1425.0 &

               \phn\phn\phn$1.51$\phn &            \phn\phn\phn$1.67$\phn &              \phn0.19 &
    \phn\phn7.7 \\
  2005 Mar 10    & 53439.4845       &       \phn1425.0 &

               \phn\phn\phn$1.41$\phn &            \phn\phn\phn$1.12$\phn &              \phn0.18 &
    \phn\phn7.7 \\
\\[0.02in]
  2005 Jan \phn4\tablenotemark{j} & 53374.6404       &       \phn4860.1 &

                            [$-0.03$] &                            \nodata &              \phn0.13 &
    \phn\phn8.4 \\
  2005 Feb 20    & 53421.5750       &       \phn4860.1 &

               \phn\phn\phn$1.34$\phn &            \phn\phn\phn$1.39$\phn &              \phn0.12 &
    \phn\phn5.7 \\
  2005 Feb 23    & 53424.5891       &       \phn4860.1 &

               \phn\phn\phn$1.59$\phn &            \phn\phn\phn$1.46$\phn &              \phn0.07 &
       \phn18.5 \\
  2005 Feb 24\tablenotemark{h}    & 53425.5089       &       \phn4860.1 &

               \phn\phn\phn$1.49$\phn &            \phn\phn\phn$1.52$\phn &              \phn0.11 &
       \phn10.9 \\
  2005 Feb 27\tablenotemark{i}    & 53428.5294       &       \phn4860.1 &

               \phn\phn\phn$1.75$\phn &            \phn\phn\phn$1.86$\phn &              \phn0.06 &
       \phn17.7 \\
  2005 Mar \phn4 & 53433.5149       &       \phn4860.1 &

               \phn\phn\phn$1.89$\phn &            \phn\phn\phn$2.01$\phn &              \phn0.10 &
    \phn\phn6.6 \\
  2005 Mar \phn6 & 53435.5242       &       \phn4860.1 &

               \phn\phn\phn$1.32$\phn &            \phn\phn\phn$1.28$\phn &              \phn0.11 &
    \phn\phn7.6 \\
  2005 Mar \phn9 & 53438.5786       &       \phn4860.1 &

               \phn\phn\phn$1.36$\phn &            \phn\phn\phn$1.22$\phn &              \phn0.10 &
    \phn\phn7.9 \\
  2005 Mar 10    & 53439.4845       &       \phn4860.1 &

               \phn\phn\phn$1.08$\phn &            \phn\phn\phn$0.86$\phn &              \phn0.11 &
    \phn\phn7.7 \\
  2005 Mar 16    & 53445.5854       &       \phn4860.1 &

               \phn\phn\phn$1.80$\phn &            \phn\phn\phn$1.73$\phn &              \phn0.07&
       \phn25.2 \\
\\[0.02in]
  2005 Feb 20    & 53421.5813       &       \phn8460.1 &

         \phn\phn\phn$0.8$\phn\phn\phn &      \phn\phn\phn$1.3$\phn\phn\phn &              \phn0.15 &
    \phn\phn5.6 \\
  2005 Feb 23    & 53424.5933       &       \phn8460.1 &

               \phn\phn\phn$1.41$\phn &            \phn\phn\phn$1.48$\phn &              \phn0.08 &
       \phn11.5 \\
  2005 Feb 27\tablenotemark{i}    & 53428.5402       &       \phn8460.1 &

               \phn\phn\phn$1.57$\phn &            \phn\phn\phn$1.56$\phn &              \phn0.05 &
       \phn15.8 \\
  2005 Mar \phn4 & 53433.5241       &       \phn8460.1 &

               \phn\phn\phn$1.53$\phn &            \phn\phn\phn$1.50$\phn &              \phn0.09 &
    \phn\phn6.3 \\
  2005 Mar \phn6 & 53435.5341       &       \phn8460.1 &

               \phn\phn\phn$1.14$\phn &            \phn\phn\phn$1.31$\phn &              \phn0.13 &
    \phn\phn6.7 \\
  2005 Mar \phn9 & 53438.5886       &       \phn8460.1 &

               \phn\phn\phn$1.10$\phn &            \phn\phn\phn$1.02$\phn &              \phn0.10 &
    \phn\phn6.7 \\
  2005 Mar 10    & 53439.5019       &       \phn8460.1 &

               \phn\phn\phn$0.80$\phn &            \phn\phn\phn$0.71$\phn &              \phn0.09 &
    \phn\phn6.1 \\
  2005 Mar 14\tablenotemark{h}    & 53443.5210       &       \phn8460.1 &

                           [\phs0.15] &                            \nodata &              \phn0.13 &
    \phn\phn5.6 \\
\\[0.02in]
  2005 Feb 27    & 53428.5294       &          22460.1 &

                           [\phs0.27] &                            \nodata &              \phn0.30&
    \phn\phn8.5 \\
  2005 Mar \phn9 & 53438.5971       &          22460.1 &

                           [\phs0.56] &                            \nodata &              \phn0.48 &
    \phn\phn3.9 \\
  2005 Mar 10    & 53439.5109       &          22460.1 &

                            [$-0.31$] &                            \nodata &              \phn0.30 &
    \phn\phn3.9 \\
  2005 Mar 16    & 53445.5871       &          22460.1 &

                               \nodata &                            \nodata &                 \nodata &
    \phn\phn3.2 \\
%---end alf.dat
%
\enddata
%
%%%%
%
\tablecomments{Radio flux densities of GRO\,J1655$-$40, as observed
  with the Very Large Array.  All observations represent continuous
  scans, and were taken in the VLA's B configuration, unless otherwise noted.
  Fast switching was employed throughout.  Phase calibrators were 
  1626$-$298 at 1425.0\,MHz; 1607$-$335 at 4860.1\,MHz and 8460.1\,MHz;
  and 1650$-$297 at 22460.1\,MHz.  The flux density scale was set by
  contemporaneous observations of 3C\,286, unless otherwise noted.
  }
%
%%%%
%
\tablenotetext{a}{UT date at mid-point of observations, before flagging}
\tablenotetext{b}{Modified Julian Day at mid-point of observations (before
  flagging): $MJD\equiv JD-2400000.5$.}
\tablenotetext{c}{Mean observing frequency. 
  % For VLA data, unless otherwise
  % noted, this
  This is the arithmetic mean of two independently-tuned 50\,MHz
  bands, observed simultaneously in both circular polarizations. 
  % The standard observing frequencies used at the VLA are given in
  % Table~\ref{tab:vlafreq}.
  }
% \tablenotetext{d}{A, B, C, D, BnA, CnB, DnC $\rightarrow$ VLA configurations;
%    A/B, B/C, C/D, D/A $\rightarrow$ VLA moving between the indicated
%    configurations;
%    NVSS $\rightarrow$ NRAO VLA Sky Survey (D configuration)}
% \tablenotetext{e}{Calibrator used to set the absolute flux density scale;
%   see Table~\ref{tab:fluxcal}.  Where none is listed, the scale was set by
%   linearly interpolating surrounding values measured for the phase
%   calibrator.}
\tablenotetext{d}{Peak flux density.  Unbracketed numbers represent the
  observed peak, after removing a planar background fit to the surrounding
  pixels; see text.  Bracketed numbers represent
  non-detections, and give the value at the known position of the source.
%  If the source position is not known to better than the synthesized beam
%  (point spread function), two numbers are given, representing the minimum
%  and maximum observed within the astrometric error circle.
  }
\tablenotetext{e}{Integrated flux density (for detections only), after
  removing a planar background fit to the neighboring pixels, as described
  in the text.  This
  represents the sum of the believable emission from the source.
  % An
  % appended $*$ indicates that the source is definitely extended; otherwise
  Any difference between this and the peak flux density reflects
  uncertainties in the images.}
\tablenotetext{f}{Root-mean-square (rms) noise in the image, as determined
  from a Gaussian fit to the distribution of flux densities in a
  region without any (known) emission.  This represents a lower limit on the
  uncertainty in the flux density, and a comparison of the peak flux density
  to this rms noise gives a reasonable signal-to-noise ratio for judging
  the statistical reliability of a detection.}
\tablenotetext{g}{Time on-source, before flagging.  Unless noted otherwise,
  this is a continuous observation, centered on the listed MJD, apart from
  interruptions for phase calibration.}
% \tablenotetext{j}{Misc.\ notes, including: fs (fast switching);
%   refptg (referenced pointing); $\phi$-cal (phase self-calibrated);
%   $A\&\phi$-cal (amplitude and phase self-calibrated); sp.line (observed in
%   spectral line mode, with half the usual bandwidth -- see text).} 
%
\tablenotetext{h}{There were no contemporaneous observations of 3C\,286
  during this run.  Instead, the flux density scale was set by linearly
  interpolating the flux density of the phase calibrator from surrounding
  observations.  This additional error thus introduced is not significant
  for the data reported here.}
\tablenotetext{i}{Observations at 1425, 4860, and 8460\,MHz on
  2005 Feb 27 were made up of two sets of scans, separated by about an
  hour.  The values reported here result from co-adding all of these data;
  the two sets of scans in all cases agree within $1\sigma$ with the
  averages reported here.}
\tablenotetext{i}{The 2005 Jan 4 data were taken in the VLA's A
  configuration.}
%%%%
%
\end{deluxetable}
%
%%%%%%%%%%%%%%%%%%%%%%%%%%%%%%%%%%%%%%%%%%%%%%%%%%%%%%%%%%%%%%%%%%%%%%%%%%%%%%%%
%

\newpage

\begin{deluxetable}{l}
\tablewidth{0pt}
\tabletypesize{\footnotesize}
\tablecaption{Spectral Fits to the GRO~J1655--40 LHS 
Combined Spectrum \tablenotemark{a}\label{model_fits}}
\tablehead{(POWER LAW + BLACK BODY)$\times$CUTOFF\tablenotemark{b}}

\startdata
$\Gamma$,\phantom{aaaaaaaaaaaaaaaaaaaaaaaaaaaaaaaaaa}   $1.35\pm0.03$ \\
$kT_{col}$, keV \phantom{aaaaaaaaaaaaaaaaaaaaaaaaaa}  $1.00_{-0.06}^{+0.06}$\\
$E_{fold}$, keV \phantom{aaaaaaaaaaaaaaaaaaaaaaaaaa} $181_{-18}^{+33}$\\
$\chi^2_\nu\, (N_{dof})$ \phantom{aaaaaaaaaaaaaaaaaaaaaaaaaaa}  1.16 (338)\\
\cutinhead{BMC$\times$CUTOFF} \\
$\Gamma$, \phantom{aaaaaaaaaaaaaaaaaaaaaaaaaaaaaaaaa}  $1.36_{-0.07}^{+0.04}$ \\
$kT_{col}$, keV \phantom{aaaaaaaaaaaaaaaaaaaaaaaaaa}  $0.71_{-0.03}^{+0.06}$\\
$f$  \phantom{aaaaaaaaaaaaaaaaaaaaaaaaaaaaaaaaaa} $0.82_{-0.05}^{+0.03}$\\
$E_{fold}$, keV  \phantom{aaaaaaaaaaaaaaaaaa.aaaaaaa} $194_{-19}^{+35}$\\
$\chi^2_\nu\, (N_{dof})$  \phantom{aaaaaaaaaaaa.aaaaaaaaaaaaaa} 1.16 (338) \\
\cutinhead{COMPTT + BLACK BODY} \\
$kT_{col}$, keV\phantom{aaaaaaaaaaaaaaaaaaaaaaaaaaa}  $0.60_{-0.04}^{+0.06}$ \\
$kT_{e}$, keV \phantom{aaaaaaaaaaaaaaaaaaaaaaaaaaa}  $37_{-3}^{+6}$\\
$\tau$  \phantom{aaaaaaaaaaaaaaaaaaaaaaaaaaaaaaaaaa} $4.4_{-0.2}^{+0.3}$\\
$\chi^2_\nu\, (N_{dof})$  \phantom{aaaaaaaaaaaa.aaaaaaaaaaaaaa} 1.23 (338) \\
\cutinhead{(PEXRAV+DISKBB)$\times$CUTOFF} \\
$\Gamma$, \phantom{aaaaaaaaaaaaaaaaaaaaaaaaaaaaaaaaa}  $1.35\pm 0.06$ \\
$R$,\phantom{aaaaaaaaaaaaaaaaaaaaaaaaaaaaaaaaaa} $ 0.12\pm0.10$ \\
$T_{in}$, keV \phantom{aaaaaaaaaaaaaaaaaaaaaaaaaaaa}  $1.49\pm0.06$\\
$E_{fold}$, keV  \phantom{aaaaaaaaaaaaaaaaaa.aaaaaaa} $196\pm 48$\\
$\chi^2_\nu\, (N_{dof})$  \phantom{aaaaaaaaaaaa.aaaaaaaaaaaaaa} 1.15 (337) \\
\enddata

\tablenotetext{a}{In addition to the listed components each fit includes also 
a narrow GAUSSIAN with the energy $~6.4$ keV to model the iron line.}
\tablenotetext{b}{For CUTOFF, the XSPEC multiplicative model HIGHECUT was used
with fixed E$_{cut}$=0.}
\end{deluxetable}

\newpage
\begin{deluxetable}{ll}
\tablewidth{0pt}
\tabletypesize{\footnotesize}
\tablecaption{Instrument Cross Calibration\tablenotemark{a}}
\tablehead{\colhead{Detector} & \colhead{$N_{DET}/N_{PCA}$} \\ }
\startdata
\cutinhead{{\it INTEGRAL}} \\
ISGRI & $0.87\pm0.03$ \\
SPI & $1.05_{-0.12}^{+0.13}$\\
JEM-X & $0.59\pm0.05$ \\
\cutinhead{{\it RXTE}} \\
HEXTE A & $0.81\pm0.02$\\
HEXTE B & $0.80\pm0.02$\\
\enddata

\tablenotetext{a}{The normalization of the incident flux 
required by each instrument is given relative to that required 
by the {\it RXTE} PCA.}
\label{cross-norm}
\end{deluxetable}

\end{document}